\documentclass[10pt,a4paper]{article}
\usepackage{amssymb}
\usepackage{amsmath}
\usepackage{amsthm}
\usepackage{latexsym}
\usepackage[dvips]{epsfig}
\usepackage{mathrsfs}
\usepackage{eufrak}
\usepackage{bm}
\usepackage{stmaryrd}
\usepackage{authblk}

\usepackage{enumerate}

\usepackage{tikz}

\theoremstyle{plain}

\newtheorem{lemma}{Lemma}
\newtheorem{theorem}{Theorem}

\def\bma{{\bm a}}
\def\bmb{{\bm b}}
\def\bmc{{\bm c}}
\def\bmd{{\bm d}}
\def\bme{{\bm e}}
\def\bmf{{\bm f}}
\def\bmg{{\bm g}}
\def\bmh{{\bm h}}
\def\bmi{{\bm i}}
\def\bmj{{\bm j}}
\def\bmk{{\bm k}}

\def\bmn{{\bm n}}

\def\bmv{{\bm v}}

\def\bmx{{\bm x}}

\def\bmzero{{\bm 0}}

\def\bmA{{\bm A}}
\def\bmB{{\bm B}}
\def\bmC{{\bm C}}
\def\bmD{{\bm D}}
\def\bmE{{\bm E}}
\def\bmF{{\bm F}}

\def\bmL{{\bm L}}

\def\bmP{{\bm P}}
\def\bmQ{{\bm Q}}

\def\bmS{{\bm S}}
\def\bmT{{\bm T}}
\def\bmX{{\bm X}}

\def\bmalpha{{\bm \alpha}}

\def\bmdelta{{\bm \delta}}

\def\bmomega{{\bm \omega}}
\def\bmlambda{{\bm \lambda}}

\def\bmsigma{{\bm \sigma}}

\def\bmGamma{{\bm \Gamma}}

\def\bmpartial{{\bm \partial}}
\def\bmnabla{{\bm \nabla}}

\newcounter{mnote}

\def\cylinderhorizon{{\mathcal{H}}}
\def\cylinderinfinity{{\mathcal{I}}}
\def\regionIIIslices{{\mathcal{R}}}
\begin{document}


\title{\textbf{On the conformal structure of the extremal
    Reissner-Nordstr\"om spacetime}}

\author[,1,2]{Christian L\"ubbe \footnote{E-mail address:{\tt c.luebbe@qmul.ac.uk}, {\tt c.luebbe@ucl.ac.uk}}}
\author[,2]{Juan Antonio Valiente Kroon \footnote{E-mail address:{\tt j.a.valiente-kroon@qmul.ac.uk}}}

\affil[1]{Department of Mathematics, University College London, Gower
  Street, London WC1E 6BT, UK}

\affil[2]{School of Mathematical Sciences, Queen Mary, University of London,
Mile End Road, London E1 4NS, United Kingdom}

\maketitle

\begin{abstract}
We analyse various conformal properties of the extremal
Reissner-Nordstr\"om spacetime. In particular, we obtain conformal
representations of the neighbourhoods of spatial infinity, timelike
infinity and the cylindrical end ---the so-called cylinders at spatial infinity and at
the horizon, respectively--- which are regular with respect to the
conformal Einstein field equations and their associated initial data
sets. We discuss possible
implications of these constructions for the propagation of test fields
and non-linear perturbations of the gravitational field close to the horizon.
\end{abstract}



\section{Introduction}

The analysis of the non-linear stability of stationary black holes is,
no doubt, one of the key open problems in contemporary mathematical
general relativity. Among stationary black hole spacetimes, extremal
ones are of particular interest. The simplest example of a extremal
black hole is given by the so-called extremal Reissner-Nordstr\"om
---a static, spherically symmetric solution to the Einstein-Maxwell
equations. 

The mainstream approach to the analysis of the non-linear stability of
black holes has consisted, in a first instance, of the  study of the
linear evolution equations (the wave equations or the Maxwell
equations) on the black hole spacetime. This approach has also been
pursued for the extremal Reissner-Nordstr\"om space time, and has
provided valuable insights ---see e.g. \cite{Are11b,Are11a,DaiDot12}. 

The conformal Einstein field equations and the conformal methods based upon them have been used with remarkable success to understand the existence and
stability of \emph{asymptotically simple spacetimes} ---see
e.g. \cite{Fri99} for a review. In view of this success, it is natural to ask whether it is
possible to adapt these ideas to analyse the stability of black hole
spacetimes. One of the underlying strategies in this conformal
approach is to obtain a detailed understanding of the geometric
structure of the background solution under consideration in order to construct 
an evolution problem which is as simple as possible. In this respect, the conformal structure of the
extremal Reissner-Nordstr\"om seems particularly amenable to a
detailed analysis. 

In \cite{LueVal13b} it has been shown that the
domain of outer communication of the extremal Reissner-Nordstr\"om
spacetime can be covered by a non-intersecting congruence of curves
with special conformal properties (the so-called \emph{conformal
  curves}). These curves are of special interest as they provide a
simple expression for a conformal factor which, in turn, could be used
to obtain a conformal representation of the spacetime. Moreover,
these curves can be used as the cornerstone of a gauge for the
conformal evolution equations. These ideas have been used in \cite{Val12}
to obtain global numerical evaluations of the extremal
Reissner-Nordstr\"om spacetime. 

One of the insights obtained from the analysis of \cite{LueVal13b} is
the special role played in the conformal geometry by the points $i^\pm$ corresponding,
respectively, to future and past timelike infinity. Recall that timelike geodesics which are not crossing the horizon start at $i^-$ and end at $i^+$. Although the analysis
of the conformal curves does not allow to conclude whether these points
 are regular points of the conformal geometry, it nevertheless shows
that the extremal Reissner-Nordstr\"om is, in a particular sense, more
regular at these points than, say, the Schwarzschild black hole or a
non-extremal Reissner-Nordstr\"om solution. This statement is justified
by the following observation: In the case of the extremal Reissner-Nordstr\"om solution, the conformal curves constructed in \cite{LueVal13b} which are passing through
$i^\pm$ remain always timelike, while in the non-extremal case the curves become null
at $i^\pm$. In view of the conformal properties of the conformal
curves in the non-extremal case, this degeneracy in their causal character indicates a
degeneracy of the conformal structure ---see also the discusion in
\cite{Fri03a}. 

A property of the extremal Reissner-Nordstr\"om solution is that is
possesses a conformal discrete isometry which, roughly speaking, maps
the domain of outer communication into itself via an inversion of the
radial coordinate and 
the horizon into null infinity ---see \cite{CouTor84}. Moreover, this
conformal inversion maps the black hole region of the
extremal Reissner-Nordstr\"om to the \emph{negative mass} extremal
Reissner-Nordstr\"om spacetime, and viceversa. A similar
property is also present in the extremal Reissner-Nordstr\"om-de
Sitter spacetime ---see \cite{Bra04}. Initially this conformal isometry was viewed as a mere curiosity, but recently it has been used in
\cite{BizFri13} to put in correspondence a class of conserved quantities on the horizon observed in \cite{Are11a,Are11b} with the so-called Newman-Penrose constants defined at null infinity
---see \cite{ExtNewPen69,NewPen65,NewPen68}. 

The purpose of this article is to exploit the discrete conformal
isometry of the extremal Reissner-Nordstr\"om spacetime to obtain a
representation of timelike infinity for which the conformal field
equations and their associated initial data are regular. In
\cite{Fri98a} it has been shown that a conformal representation with
these properties can be obtained for the spatial infinity of a vacuum
spacetime with time-reflection symmetry ---the so-called
\emph{cylinder at spatial infinity} $\cylinderinfinity$. In this article,
it is shown that a similar construction can be implemented for the
spatial infinity of the extremal Reissner-Nordstr\"om spacetime ---the
analysis being independent of the sign of the mass. By applying the
conformal inversion to this construction we obtain the \emph{cylinder
at timelike infinity} $\bar{\cylinderinfinity}_{i^+}$, a representation of the
neighbourhood of the point $i^+$ for which the conformal field
equations and their data are regular. Remarkably, this strategy can
also be used to obtain a similar representation of a neighbourhood of
the cylindrical end $c^0$ of the extremal
Reissner-Nordstr\"om spacetime ---the \emph{cylinder at the singularity}
$\cylinderhorizon_{c^0}$. We call these constructions collectively
the \emph{cylinders at the horizon} $\cylinderhorizon$.

The cylinders at the horizon inherit, in a natural manner, many of the
properties of the cylinder at spatial infinity. Crucially, they are 
\emph{total characteristics} of the evolution equations implied by the
conformal field equations ---that is, the evolution equations
completely reduce to an interior system of transport equations. As a
consequence, no boundary conditions can be specified at the cylinders. The
solution to the evolution equations is fully determined by its value
at some section of the cylinder. A further analogy between the
cylinder $\cylinderhorizon$ at the horizon and the cylinder ${\cylinderinfinity}$ at spatial infinity is given by the following:
The cylinders terminate in sets with the topology of $\mathbb{S}^2$
where one observes a degeneracy of the symmetric hyperbolic evolution
equations deduced from spinorial field equations whose principal part
is of the form
\[
\nabla^{AA'} \phi_{AB\cdots Q} , \qquad \phi_{AB\cdots Q} =
\phi_{(AB\cdots Q)}.
\]
In the case of the the cylinder at spatial infinity, it
has been possible to relate this
degeneracy in the evolutions equations with the appearance of
potential \emph{obstructions to the smoothness of null infinity}
---see \cite{Fri98a,Val03a,Val04a,Val04d,Val07b}. It is conjectured
that similar obstructions will arise from the analysis
of transport equations at the cylinders at the horizon. This question
and its potential implications for the analysis of non-linear
perturbations of the extremal Reissner-Nordstr\"om spacetime will be
analysed elsewhere.  

There is, however, a crucial difference between the cylinder ${\cylinderinfinity}$ at
spatial infinity and the cylinder $\cylinderhorizon$ at the horizon. While the
conformal factor associated to the former representation vanishes at
$\cylinderinfinity$, in the later representations the respective conformal
factors do not vanish at $\cylinderhorizon$. Thus,
strictly speaking, the cylinders at the horizon are not part of
the conformal boundary. 

\medskip
A natural question to be asked at this point is how crucial is the
conformal isometry of the extremal Reissner-Nordstr\"om in the
construction presented in this article. While it is of great
value in order to gain intuition about the underlying structures, we claim it is not
essential. Once the key aspects of the construction have been identified,
the results of this paper could have been obtained without using this
isometry at the expense of lengthier arguments. This claim suggests
the possibility of performing a similar analysis in other extremal
black hole spacetimes, in particular the \emph{extremal Kerr solution}.

\subsubsection*{Outline of the article}
In Section \ref{Section:BasicExpressions} we provide a summary of some
basic facts concerning the extremal Reissner-Nordstr\"om spacetime
(various types of coordinates, Penrose diagrams, properties of the
conformal isometry) which will be used throughout this
article. Section \ref{Section:ConformalGeodesics} discusses the basic
properties of conformal geodesics in electrovacuum spacetimes which
will be used in our analysis. Section
\ref{Section:TimeSymmetricSlices} provides an analysis of the
conformal properties of time symmetric hypersurfaces in the extremal
Reissner-Nordstr\"om spacetime and the initial data for the conformal Einstein
field equations. This analysis is key to identify the singular
behaviour of various conformal fields at timelike infinity and the
cylindrical asymptotic end. Section
\ref{Section:CylinderSpatialInfinity} discusses the construction of
the cylinder at spatial infinity for the extremal Reissner-Nordstr\"om
spacetime. This construction is used, in turn, in Section
\ref{Section:CylindersHorizon} to motivate and implement the representation of the
cylinders at the horizon. Conclusions and possible implications of the
present analysis to the propagation of fields close to timelike
infinity and the cylindrical asymptotic end are discussed in Section
\ref{Section:ConcludingRemarks}. Finally, an appendix provides some details
about the transformation formulae relating objects in the cylinders at
spatial infinity and at the horizon. 

\subsubsection*{Notation}
Our signature convention for spacetime (Lorentzian) metrics is
$(+---)$. In what follows ${}_a,\, {}_b, \, {}_c, \ldots$ denote
spacetime tensorial indices while ${}_\bma,\, {}_\bmb, \, {}_\bmc,
\ldots$ correspond to spacetime frame indices taking the values
$0,\ldots,3$. Spatial tensorial indices will be denoted by ${}_i, \,
{}_j,\, {}_k, \ldots$ while spatial frame indices by ${}_\bmi, \,
{}_\bmj, \, {}_\bmk$. Part of analysis will require the use of
spinors. In this respect we make use of the general conventions of
Penrose \& Rindler \cite{PenRin84}. In particular, ${}_A, \,{}_B,
\,{}_C, \dots$ denote abstract spinorial indices, while  ${}_\bmA, \,{}_\bmB,
\,{}_\bmC, \dots$ indicate frame spinorial indices with respect to
some specified spin dyad $\{ \delta_\bmA \}$. 

Index-free notation will also be used in many places. Given a 1-form $\bmomega$, its
pairing with a vector $\bmv$ will be denoted by $\langle \bmomega,
\bmv \rangle$. Given a metric $\bmg$, its contravariant counterpart
will be denoted by $\bmg^\sharp$. The operation of raising the
index of the 1-form $\bmomega$ will be denoted by $\bmomega^\sharp
\equiv \bmg^\sharp(\bmomega, \cdot)$. Similarly, the lowering of the
index of the vector $\bmv$ will be denoted by $\bmv^\flat \equiv
\bmg(\bmv,\cdot)$. Given a connection $\bmnabla$, the covariant
directional derivative along a curve with tangent $\bmv$ will be
denoted by $\nabla_\bmv$. 

Various connections will be used throughout. The connection
$\tilde{\bmnabla}$ will always denote the Levi-Civita connection of a
Lorentzian metric $\tilde{\bmg}$ satisfying the Einstein-Maxwell field
equations ---hence, we call it the \emph{physical
  connection}. Connections conformally related to $\tilde{\bmnabla}$
will be denoted by $\bmnabla$ and $\check{\nabla}$ and will be called
\emph{unphysical}. Finally, $\hat{\bmnabla}$ will denote a Weyl
connection in the conformal class of $\tilde{\bmnabla}$.

\section{Basic expressions}
\label{Section:BasicExpressions}
The \emph{extremal Reissner-Nordstr\"om spacetime} is the solution to the
Einstein-Maxwell field equations
\begin{subequations}
\begin{eqnarray}
&& \tilde{R}_{ab} -\tfrac{1}{2}\tilde{R}\, \tilde{g}_{ab}  =
\tilde{F}_{ac}\tilde{F}^c{}_b
-\tfrac{1}{4}\tilde{g}_{ab}
\tilde{F}_{cd}\tilde{F}^{cd}, \label{EMFE1}\\
&& \tilde{\nabla}^b \tilde{F}_{ab}=0, \label{EMFE2}\\
&& \tilde{\nabla}_{[a} \tilde{F}_{bc]}=0, \label{EMFE3}
\end{eqnarray}
\end{subequations}
given in \emph{standard spherical coordinates} $(t,\tilde{r},\theta,\varphi)$ by
\begin{subequations}
\begin{eqnarray}
&& \hspace{-10mm}{\tilde{\bm g}} = \left(
  1-\frac{m}{\tilde{r}}\right)^2 \mathbf{d}t\otimes \mathbf{d}t -\left( 1-\frac{m}{\tilde{r}}\right)^{-2} \mathbf{d} \tilde{r}\otimes \mathbf{d} \tilde{r}
- r^2 {\bm\sigma} \label{StandardRN}\\
&& \hspace{-10mm} \tilde{\bm F} =
\pm \frac{m}{2\tilde{r}^2} \mathbf{d}t \wedge \mathbf{d}\tilde{r}. \label{FaradayRN}
\end{eqnarray}
\end{subequations}
where 
\[
{\bm \sigma} \equiv \left(\mathbf{d}
  \theta \otimes \mathbf{d} \theta + \sin^2 \theta \mathbf{d}\varphi \otimes \mathbf{d} \varphi \right)
\]
is the standard metric of $\mathbb{S}^2$. The discussion in the this
article will be concerned with both $m>0$ and $m<0$ cases of the
solution \eqref{StandardRN}-\eqref{FaradayRN}.  For the sake of
generality in our analysis, we will use the general notation $r$, $m$,
$\tilde{\bm g}$, etc. so that both cases can be discussed
simultaneously. However, where it is necessary to avoid confusion or
ambiguity, we distinguish the two mass cases by the corresponding
subscript, i.e. on the positive Reissner-Nordstr\"om solution we use
$r_+$, $m_+$, $\tilde{\bm g}_+$, etc. and on the negative mass
spacetime we use $r_-$, $m_-$, $\tilde{\bm g}_-$, etc. In line with
the above, we will denote the maximal analytic extension in the
positive mass case ($m>0$) by
$(\widetilde{\mathcal{M}}_+,\tilde{\bmg}_+)$ (see e.g. Carter
\cite{Car66b,Car73a}).  Similarly, the maximal analytic extension in
the negative mass case ($m<0$) will be denoted by
$(\widetilde{\mathcal{M}}_-,\tilde{\bmg}_-)$.

\subsection{Isotropic and null coordinates}
In this article it will be more convenient to make
use of the \emph{isotropic
  coordinate} $r$ defined by
\[
r=\tilde{r}-m, \qquad \tilde{r}=r+m.
\]
In terms of this coordinate the metric of the extreme
Reissner-Nordstr\"om spacetime is given by
\begin{equation}
\tilde{\bmg} = \left(1+\frac{m}{r}\right)^{-2}\mathbf{d}t \otimes
\mathbf{d}t -\left( 1+\frac{m}{r} \right)^2\left( \mathbf{d}r\otimes
  \mathbf{d}r + r^2 {\bm \sigma}\right).
\label{eRNIsotropic}
\end{equation}
In the positive mass case, this metric is well defined for both
positive and negative values of the coordinate $r$: the region for
which $r>0$ corresponds to the domain of outer communication of a
static black hole spacetime while that for which $r<0$ corresponds to
the black hole region.  In the negative mass case there are no
horizons, and $r = r_- \in [|m|, \infty)$. The vector $\bmpartial_t$ is clearly a Killing vector for
the metric \eqref{eRNIsotropic}. Except at the horizon (in the
positive mass case) it is always timelike ---by contrast to the
analogous vector field in the Schwarzschild spacetime or the non-extremal Reissner-Nordstr\"om  spacetime.

\medskip
Retarded and advanced null coordinates can be introduced via 
\[
u = t -\left( r -\frac{m^2}{r} + 2m \mbox{ln} |r|
\right), \qquad v = t +\left( r -\frac{m^2}{r} + 2m \mbox{ln} |r|
\right), 
\]
to obtain the line elements
\begin{subequations}
\begin{eqnarray}
&& \tilde{\bmg}(u,r) = \frac{r^2}{(r+m)^2} \mathbf{d} u \otimes \mathbf{d} u +
(\mathbf{d} u \otimes \mathbf{d}r + \mathbf{d}r\otimes
\mathbf{d}u ) - (r+m)^2\bmsigma, \label{RetardedTimeMetric} \\
&&  \tilde{\bmg}(v,r) = \frac{r^2}{(r+m)^2}\mathbf{d} v \otimes \mathbf{d} v -
(\mathbf{d} v \otimes \mathbf{d}r + \mathbf{d}r\otimes
\mathbf{d}v ) - (r+m)^2 \bmsigma, \label{AdvancedTimeMetric}
\end{eqnarray}
\end{subequations}
where in the previous expressions the notation $\tilde{\bmg}(u,r)$ and
$\tilde{\bmg}(v,r)$ is intended to highlight the particular choice
of coordinates being used to express the metric
$\tilde{\bmg}$. Similarly, in \eqref{StandardRN} and \eqref{eRNIsotropic} one could have written
$\tilde{\bmg}(t,\tilde{r})$ and $\tilde{\bmg}(t,r)$, respectively. 

\subsection{Penrose diagrams}
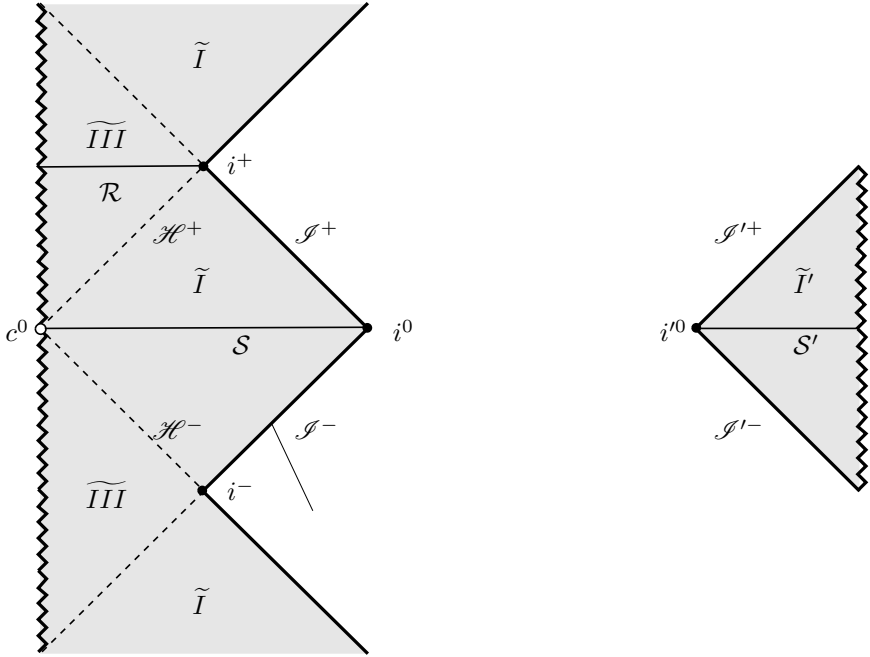
\begin{figure}[t]
\begin{center}
\includegraphics[width=0.8\textwidth]{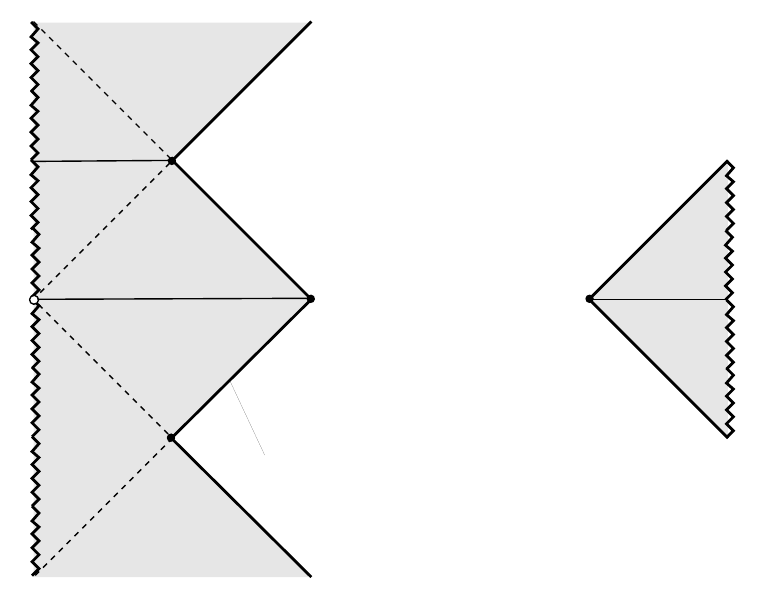}
\put(-190,127){$i^0$}
\put(-90,127){$i^{\prime 0}$}
\put(-335,127){$c^0$}
\put(-227,90){$\mathscr{I}^-$}
\put(-70,90){$\mathscr{I}^{\prime -}$}
\put(-70,164){$\mathscr{I}^{\prime +}$}
\put(-227,164){$\mathscr{I}^+$}
\put(-252,190){$i^+$}
\put(-252,67){$i^-$}
\put(-280,164){$\mathscr{H}^+$}
\put(-280,90){$\mathscr{H}^-$}
\put(-265,145){$\widetilde{I}$}
\put(-265,230){$\widetilde{I}$}
\put(-265,24){$\widetilde{I}$}
\put(-40,145){$\widetilde{I}'$}
\put(-305,200){$\widetilde{III}$}
\put(-305,64){$\widetilde{III}$}
\put(-40,122){$\mathcal{S}'$}
\put(-250,122){$\mathcal{S}$}
\put(-300,180){$\regionIIIslices$}
\end{center}
\caption{Conformal diagram of the extremal Reissner-Nordstr\"om
spacetime in the positive mass (left) and negative mass (right)
cases. The definitions of the regions $\widetilde{I}$,
$\widetilde{III}$ and $\widetilde{I}'$ are given in the main
text. The future and past null infinities of the positive mass case are denoted,
respectively, by $\mathscr{I}^\pm$, the horizons by $\mathscr{H}^\pm$,
the future and past timelike infinities by $i^\pm$, the spatial
infinities by $i^0$ and the cylindrical end by $c^0$. A similar
notation is used for the analogous sets  in the negative
mass case. The
hypersurfaces $\mathcal{S}$ and $\regionIIIslices$ are time symmetric
slices in the positive mass case, while $\bar{\mathcal{S}}'$ is a time symmetric
slice in the negative mass case  ---see Section
\ref{Section:TimeSymmetricSlices}.}
\label{ConformalDiagrameRN}
\end{figure}

The Penrose diagrams of the extremal Reissner-Nordstr\"om solutions are
well-known. They were first discussed in \cite{Car66b} ---see also
\cite{GriPod09}. The diagrams for both
the positive and negative mass cases are given in given in Figure
\ref{ConformalDiagrameRN} for reference. 

For the purposes of the
subsequent discussion it is convenient to identity the following
subsets of the positive mass extremal Reissner-Nordstr\"om spacetime:
\begin{eqnarray*}
&& \widetilde{I} \equiv \{ p \in \widetilde{\mathcal{M}}_+ \; | \; -\infty < v(p) < \infty, \; 0< r(p)<\infty \}, \\\
&& \widetilde{III} \equiv  \{ p \in \widetilde{\mathcal{M}}_+ \; | \; -\infty < v(p) < \infty, \; -m < r(p) <0 \}, \\
&& \mathscr{H}^+ \equiv \{ p \in \widetilde{\mathcal{M}}_+ \; | \;
-\infty < v(p) < \infty , \; r(p)=0  \}, \\
&& \mathscr{H}^- \equiv \{ p \in \widetilde{\mathcal{M}}_+ \; | \;
-\infty < u(p) < \infty , \; r(p)=0  \},
\end{eqnarray*}
describing, respectively, the domain of outer communication, the black
hole region and the future and past horizons. In the negative mass
case there is no black hole region so that one has only
\[
\widetilde{I}' \equiv \{ p \in \widetilde{\mathcal{M}}_- \; | \; -\infty < v(p) < \infty, \; |m|< r(p)<\infty \}.
\]

\medskip
\noindent
\textbf{Remark.} The notation of $\widetilde{I}$ and $\widetilde{I}'$
was chosen to indicate their physical resemblence. Both regions
describe the asymptotic region of the black hole, however
$\widetilde{I}$ borders the horizon while $\widetilde{I}'$ borders a
naked singularity. In fact, as we will see below, in terms of its
conformal geometry $\widetilde{I}'$ is more alike that of $\widetilde{III}$.

\subsection{The discrete conformal isometry of the extremal Reissner-Nordstr\"om spacetime}
\label{Section:ConformalInversion}

In \cite{CouTor84} it has been shown that the extremal Reissner-Nordstr\"om
possesses a \emph{conformal discrete isometry} ---see also
\cite{BizFri13} for a more elaborated discussion. This isometry is best
expressed in isotropic coordinates and it is implemented by the
inversion of the radial isotropic coordinate $\iota: r
\rightarrow m^2/r$, so $\iota$ is an involution. Let $\iota_*$ denote the push-forward map implied
by $\iota$. As $\iota$ is an involution for $r\neq 0$, its action well defined on both
covariant and contravariant tensors. 

In order to discuss the effect of the discrete conformal isometry it
is convenient to introduce the \emph{inversion radial coordinate}
$\varrho\equiv m^2/r$, so that $\iota: r \to \varrho$ and $\iota:\varrho \to r$. Moreover, it can be verified that the conformal inversion interchanges
the null coordinates $u$ and $v$ ---that is, one has that
\[
\iota_* u \equiv u \circ \iota = v, \qquad \iota_* v \equiv v \circ \iota =u.
\]
A direct computation using the metric $\tilde{\bmg}$ in the form \eqref{eRNIsotropic}shows that
\[
\tilde{\bmg}(t,\varrho) = \left( \frac{m}{\varrho}\right)^2\left(\frac{\varrho^2}{(\varrho+m)^2}\mathbf{d}t\otimes \mathbf{d}t -(\varrho+m)^2 (\mathbf{d}\varrho\otimes\mathbf{d} \varrho +\varrho^2 \bmsigma)  \right).
\]
Thus, the inversion $r\mapsto m^2/r$ is implemented in the above metric by simply making the replacement $\varrho \mapsto r$. Moreover
\begin{equation}
\iota_* \tilde{\bmg} = \Omega^2 \tilde{\bmg}, \qquad \mbox{with} \qquad\Omega\equiv \frac{m}{r} = \frac{\varrho}{m}.
\label{Conformomorphism}
\end{equation}
A similar computation using the expressions of the extremal Reissner-Nordstr\"om
metric in terms of the retarded and advanced coordinates $u$ and $v$,
equations \eqref{RetardedTimeMetric}-\eqref{AdvancedTimeMetric} shows
that
\[
\iota_* \tilde{\bmg}(v,\varrho) = \Omega^2 \tilde{\bmg}(u,r), \qquad \iota_* \tilde{\bmg}(u,\varrho) = \Omega^2 \tilde{\bmg}(v,r).
\]
For future use, it is convenient to define the \emph{unphysical}
metric
\begin{equation}
\check{\bmg} \equiv \Omega^2 \tilde{\bmg}, \qquad\Omega\equiv \frac{m}{r} = \frac{\varrho}{m},
\label{ConformalMetricIsometry}
\end{equation}
so that one can write
\[
\iota_* \tilde{\bmg} = \check{\bmg} \qquad \mbox{and} \qquad \iota_*  \check{\bmg} = \tilde{\bmg}  .
\]

\subsubsection*{Effect of the isometry on region $\widetilde{I}$}

From the discussion in the previous paragraphs it follows that $(I,\check{\bmg}_+(u,\varrho))$ with
\begin{eqnarray*}
&& I \equiv \{ p \in \mathbb{R} \times (0,\infty)\times \mathbb{S}^2 \;|\;
-\infty < u(p) <\infty, \; 0\leq \varrho(p)< \infty \}, \\
&& \check{\bmg}_+(u,\varrho) = \frac{\varrho^2}{(\varrho+m)^2} \mathbf{d}u\otimes \mathbf{d} u -
2(\mathbf{d}u \otimes \mathbf{d}\varrho + \mathbf{d}\varrho \otimes
\mathbf{d}u) - (\varrho+m)^2\bmsigma,
\end{eqnarray*}
is a conformal extension of $(\widetilde{I},\tilde{\bmg}_+(u,r))$ ---in
particular, $\check{\bmg}_+(u,\varrho)$ is real analytic on $I$. The
null hypersurface 
\[
\mathscr{I}^+=\{ p \in I \;|\; \varrho(p) =0  \}, 
\]
on which $\Omega=0$, $\mathbf{d}\Omega\neq 0$ represents the future
null infinity of the region $\widetilde{I}$. From the discussion in
the previous paragraphs it follows that $\iota:
\widetilde{I} \rightarrow \widetilde{I}$ can, in fact, be extended 
to a real analytic isometry 
\[
\iota : I \longrightarrow \tilde{I} \cup \mathscr{H}^+.
\]
In particular, one has that
\[
\iota( \mathscr{I}^+ )= \mathscr{H}^+. 
\]
That is, the conformal isometry $\iota$ sends future null infinity
into the future horizon (and viceversa).

\subsubsection*{Effect of the isometry on $\widetilde{\mathcal{M}}_-$}
Key for the purposes of this article is the effect of $\iota$ on the
negative mass extremal Reissner-Nordstr\"om spacetime.
In analogy to the positive mass case, one can use the conformal metric
\eqref{ConformalMetricIsometry} to construct a conformal extension of
$(\widetilde{\mathcal{M}}_-,\tilde{\bmg}_-(u,r))$, namely
$(\mathcal{M}_-,\check{\bmg}_-(u,\varrho))$ with
\begin{eqnarray*}
&&  \mathcal{M}_-\equiv \{ p \in \mathbb{R} \times (0,\infty)\times \mathbb{S}^2 \;|\;
-\infty < u(p) <\infty, \; 0\leq \varrho(p)< |m| \}, \\
&& \check{\bmg}_-(u,\varrho) = \frac{\varrho^2}{(\varrho+m)^2} \mathbf{d} u \otimes
\mathbf{d}u -2 (\mathbf{d}u \otimes
\mathbf{d}\varrho + \mathbf{d}\varrho \otimes \mathbf{d}u) - (\varrho+m)^2 \bmsigma.
\end{eqnarray*}
Due to the sign of the mass $\Omega = m/r =  -|m|/r_- < 0$. Observe
that the same happens in $\widetilde{III} $, since $r_+ \in (-|m|,
0)$. We note here that the negative sign of the conformal factor is of
no concern to us as this commonly happens when formulating the
conformal Einstein field equations across null infinity\footnote{The same happens when one conformally extends positive Reissner-Nordstr\"om across null infinity in terms of coordinates.}.

The null infinity of $\widetilde{\mathcal{M}}_-$ is given the set
\[
\mathscr{I}^{\prime +} \equiv \{ p \in \mathcal{M}_- \,|\, \varrho(p)=0  \},
\]
on which $\Omega = \varrho/m =0$,
$\mathbf{d}\Omega\neq 0$ and which consists of two disjoint null hypersurfaces. 

\medskip
Applying the discrete conformal
isometry to $\check{\bmg}_-$ one finds that 
\[
\iota_*\check{\bmg}_-(u,\varrho) = \frac{r^2}{(r+m)^2} \mathbf{d} v \otimes
\mathbf{d}v -2 (\mathbf{d}v \otimes
\mathbf{d}r + \mathbf{d}r \otimes \mathbf{d}v) - (r+m)^2 \bmsigma. 
\]
Hence, making the replacements $\bar{m}=-m$, $\bar{r}=-r$ and $\bar{v}=v$, one finds
that $\iota_*\check{\bmg}_-$ can be identified with a positive mass
extremal Reissner-Nordstr\"om metric in region $\widetilde{III}$.
Consequently, in the negative mass case, the conformal inversion can be
regarded as real analytic conformal isometry
\[
\iota: \mathcal{M}_- \longrightarrow \widetilde{III} \cup \mathscr{H}^+ \qquad \mbox{with} \qquad
\iota: \tilde{\bmg}_- \to \check{\bmg}_- = \tilde{\bmg}_+
\]
such that
\[
\iota(\mathscr{I}^{\prime+}) = \mathscr{H}^+.
\]

\section{Conformal geodesics in electrovacuum spacetimes}
\label{Section:ConformalGeodesics}
In the sequel we will make use of conformal geodesics to probe the
properties of the extremal Reissner-Nordstr\"om spacetime. Following \cite{Fri95,Fri03c,FriSch87}, a conformal geodesic is a pair $(\bmx(\tau),\tilde{\bmb}(\tau))$ 
--- consisting  of a curve $\bmx(\tau)$ in a spacetime $(\tilde{\mathcal{M}},\tilde{\bmg})$ with
parameter $\tau$ and a 1-form $\tilde{\bmb}(\tau)$ along the curve --- that
satisfies, in index-free notation, the equations
\begin{subequations}
\begin{eqnarray}
&& \tilde{\nabla}_{\dot{\bmx}} \dot{\bmx} = -2 \langle
\tilde{\bmb},\dot{\bmx}\rangle \dot{\bmx} +
\tilde{\bmg}(\dot{\bmx},\dot{\bmx}) \tilde{\bmb}^\sharp, \label{ConformalGeodesicEquation1}\\
&& \tilde{\nabla}_{\dot{\bmx}} \tilde{\bmb} = \langle \tilde{\bmb}, \dot{\bmx} \rangle
\tilde{\bmb} - \tfrac{1}{2}\tilde{\bmg}^\sharp (\tilde{\bmb},\tilde{\bmb}) \dot{\bmx}^\flat + \tilde{\bmL}(\dot{\bmx},\cdot).  \label{ConformalGeodesicEquation2}
\end{eqnarray}
\end{subequations}
Here $\dot{\phantom{x}}$ denotes differentiation with respect to the
parameter $\tau$ and $\tilde{\bmL}$ denotes the Schouten tensor of the
metric $\tilde{\bmg}$. Given a conformal geodesic
$(\bmx(\tau),\tilde{\bmb}(\tau))$, it is convenient to consider a
frame $\{ \bme_\bma \}$ propagated along curve according to the
equation
\begin{equation}
\tilde{\nabla}_{\dot{\bmx}} \bme_\bma = -\langle \tilde{\bmb},\bme_\bma\rangle
\dot{\bmx} - \langle \tilde{\bmb}, \dot{\bmx}\rangle \bme_\bma + \tilde{\bmg}
(\bme_\bma,\dot{\bmx}) \tilde{\bmb}^\sharp.
\label{WeylPropagationEquation}
\end{equation}
Frames satisfying this equation are said to be \emph{Weyl propagated}
along the conformal geodesic. In the sequel we will only use
frames for which $\bme_\bmzero =\dot{\bmx}$. 

\subsection{Canonical conformal factors}
\label{Section:CanonicalConformalFactor}

Timelike conformal geodesics allow one to single out a \emph{canonical
  representative} of the conformal class $[\tilde{\bmg}]$. To see
this, let 
\[
\bmg = \Theta^2 \tilde{\bmg}
\]
and require that
\[
\bmg(\dot{\bmx},\dot{\bmx}) = \Theta^2 \tilde{\bmg}(\dot{\bmx},\dot{\bmx})=1.
\]
By repeatedly differentiating this last equation with respect to $\tau$ and
using the conformal geodesic equations
\eqref{ConformalGeodesicEquation1}-\eqref{ConformalGeodesicEquation2}
one finds that the following equations hold along a given conformal geodesic:
\begin{subequations}
\begin{eqnarray}
&& \dot{\Theta} = \langle \tilde{\bmb}, \dot{\bmx} \rangle  \Theta, \label{ThetaEvolutionEquation1}\\
&& \ddot{\Theta} = \tfrac{1}{2} \tilde{\bmg}^\sharp (\tilde{\bmb},\tilde{\bmb})
\Theta^{-1} + \Theta \tilde{\bmL}(\dot{\bmx},\dot{\bmx}), \label{ThetaEvolutionEquation2}\\
&& \dddot{\Theta} = \Theta \tilde{\nabla}_{\dot{\bmx}} \big(
\tilde{\bmL}(\dot{\bmx},\dot{\bmx})\big) + \dot{\Theta}
\tilde{\bmL}(\dot{\bmx},\dot{\bmx}) + \tilde{\bmL}(\dot{\bmx},
\tilde{\bmb}^\sharp) \Theta^{-1}. \label{ThetaEvolutionEquation3}
\end{eqnarray}
\end{subequations}
Moreover, one also has that
\begin{subequations}
\begin{eqnarray}
&& \tilde{\nabla}_{\dot{\bmx}} \big( \Theta \langle \tilde{\bmb}, \bme_\bma
\rangle \big) = \Theta \tilde{\bmL}(\dot{\bmx},\bme_\bma) + \tfrac{1}{2}
\Theta \tilde{\bmg}^\sharp (\tilde{\bmb},\tilde{\bmb}) \tilde{\bmg}(\dot{\bmx},\bme_\bma), \label{AuxiliaryConformalGeodesics1}\\
&& \tilde{\nabla}_{\dot{\bmx}} \big( \bmg(\bme_\bma,\bme_\bmb)
\big)=0. \label{AuxiliaryConformalGeodesics2}
\end{eqnarray}
\end{subequations}
Hence, if the frame $\{ \bme_\bma\}$ is initially orthogonal, it remains
orthogonal along the whole of the conformal geodesic. In what follows,
it will be convenient to introduce the 1-form
\[
\bmd \equiv \Theta \bmb.
\]
Using equation \eqref{ThetaEvolutionEquation1} it readily follows that
\begin{equation}
d_0 = \dot{\Theta}.
\label{Equation:d0}
\end{equation}
That is, the time component of the 1-form $\bmd$ is known if the
conformal factor $\Theta$ is known. Moreover, from equation
\eqref{AuxiliaryConformalGeodesics1} one concludes that
\begin{equation}
\dot{d}_\bmi = \Theta \tilde{L}_{\bmzero\bmi},
\label{AuxiliaryConformalGeodesics3}
\end{equation}
where $d_\bmi\equiv \langle \bmd, \bme_\bmi\rangle$ and
$\tilde{L}_{\bmzero\bmi}\equiv \tilde{L}(\dot{\bmx},\bme_i)$. 

\medskip
If the spacetime satisfies the vacuum field equation
$\bm{Ric}[\tilde{\bmg}]= \lambda \tilde{\bmg}$, it follows readily
from equation \eqref{ThetaEvolutionEquation3} that $\dddot{\Theta}=0$ and one can
obtain an explicit expression for the conformal
factor along the conformal geodesic which is  quadratic in $\tau$
---see e.g. \cite{Fri95,Fri03c,FriSch87}. In the presence of matter this
simple expression for the conformal factor is no longer available, but
one can still make use of equation \eqref{ThetaEvolutionEquation3} to
evolve $\Theta$. For electrovacuum spacetimes with vanishing Cosmological constant it
follows readily from the Einstein field equations that $\tilde{\bmL} =
\tfrac{1}{2}\tilde{\bmT}$, where $\tilde{\bmT}$ is the \emph{physical
  energy-momentum tensor of the Maxwell field}. In the sequel, we will
work mostly in the conformally extended spacetime $(\mathcal{M},\bmg)$
and, accordingly, it is convenient to introduce the \emph{unphysical
  energy-momentum tensor} $\bmT \equiv \Theta^{-2} \tilde{\bmT}$, \cite{Fri91}. It
follows from the previous discussion that
\[
\tilde{\bmL} = \tfrac{1}{2} \Theta^2 \bmT.
\] 
Defining the \emph{unphysical energy density} by
$\mu\equiv\bmT(\dot{\bmx},\dot{\bmx})$, one can recast equation
\eqref{ThetaEvolutionEquation3} in the more suggestive form
\begin{equation}
\dddot{\Theta} = \tfrac{1}{2} \Theta^3 \dot{\mu} + \tfrac{3}{2}\Theta^2
\dot{\Theta} \mu + \tfrac{1}{2}\bmT(\dot{\bmx},
\tilde{\bmg}^\sharp(\bmd, \cdot)).
\label{ThetaEvolutionEquation3ALT}
\end{equation}
Thus, the evolution of the conformal factor is coupled to that of the
matter content of the spacetime. Similarly, defining the components of \emph{unphysical
  electromagnetic flux vector by} $j_\bmi \equiv
\bmT(\dot{\bmx},\bme_\bmi)$, one  finds from
\eqref{AuxiliaryConformalGeodesics3} that
\begin{equation}
\dot{d}_\bmi =
\tfrac{1}{2}\Theta^3 j_\bmi. 
\label{Equation:di}
\end{equation}
Initial data for equation \eqref{ThetaEvolutionEquation3ALT}  is
constrained by equations
\eqref{ThetaEvolutionEquation1}-\eqref{ThetaEvolutionEquation2}. One
readily has that
\begin{subequations}
\begin{eqnarray}
&& \dot{\Theta}_\star = \langle \bmd,\dot{\bmx}\rangle_\star, \label{ThetaEvolutionData1}\\
&& \Theta_\star \ddot{\Theta}_\star =
\tfrac{1}{2}\bmh^\sharp(\bmd_\star,\bmd_\star) + \Theta^4_\star \mu_\star, \label{ThetaEvolutionData2}
\end{eqnarray}
\end{subequations}
where the subindex ${}_\star$ indicates the value of the relevant
quantities on a fiduciary hypersurface $\tilde{\mathcal{S}}$.

\subsection{Behaviour under the conformal isometry}
\label{Section:CGIsometry}
In this section we analyse the behaviour of the conformal geodesic
equations and their solutions under the conformal isometry \eqref{Conformomorphism}. If
$(\bmx(\tau), \tilde{\bmb}(\tau))$ is a solution to equations
\eqref{ConformalGeodesicEquation1}-\eqref{ConformalGeodesicEquation2},
it follows readily from $\iota_* \tilde{\bmg} = \check{\bmg}$ and general
properties of isometries that
$(\iota_* \bmx (\tau), \iota_* \bmb(\tau))$ is a solution to
\begin{eqnarray*}
&& \check{\nabla}_{\iota_*\dot{\bmx}} \iota_*\dot{\bmx} = -2 \langle
\iota_*\tilde{\bmb},\iota_*\dot{\bmx}\rangle \iota_*\dot{\bmx} +
\check{\bmg}(\iota_*\dot{\bmx},\iota_*\dot{\bmx}) \iota_*\tilde{\bmb}^\sharp, \\
&& \check{\nabla}_{\iota_*\dot{\bmx}} \iota_*\tilde{\bmb} = \langle \iota_*\tilde{\bmb}, \iota_*\dot{\bmx} \rangle
\iota_*\tilde{\bmb} - \tfrac{1}{2}\check{\bmg}^\sharp
(\iota_*\tilde{\bmb},\iota_*\tilde{\bmb}) \iota_*\dot{\bmx}^\flat +
\check{\bmL}(\iota_*\dot{\bmx},\cdot), 
\end{eqnarray*}
---the conformal geodesic equations with respect to the connection
$\check{\bmnabla}$. Finally, observing that $\check{\bmg} = \Omega^2 \tilde{\bmg}$, and the
properties of conformal geodesics under changes of connection ---see
\cite{Fri95,Fri03c} for details--- one concludes that $(\iota_*
\bmx(\tau), \bmb(\tau))$ with 
\begin{equation}
\bmb(\tau) \equiv \iota_*
\tilde{\bmb} + \Omega^{-1} \mathbf{d}\Omega
\label{GCIsometryAction}
\end{equation}
 is a solution to the
$\tilde{\bmnabla}$-geodesic equations
\begin{eqnarray*}
&& \tilde{\nabla}_{\iota_*\dot{\bmx}} \iota_*\dot{\bmx} = -2 \langle
\bmb,\iota_*\dot{\bmx}\rangle \iota_*\dot{\bmx} +
\tilde{\bmg}(\iota_*\dot{\bmx},\iota_*\dot{\bmx}) \bmb^\sharp, \\
&& \tilde{\nabla}_{\iota_*\dot{\bmx}}\bmb = \langle \bmb, \iota_*\dot{\bmx} \rangle
\bmb - \tfrac{1}{2}\tilde{\bmg}^\sharp
(\bmb,\bmb) \iota_*\dot{\bmx}^\flat +
\tilde{\bmL}(\iota_*\dot{\bmx},\cdot). 
\end{eqnarray*}

Summarising, one has the following:

\begin{lemma}
Conformal isometries map conformal
geodesics into conformal geodesics. 
\end{lemma}

\subsection{Conformal geodesics and Weyl connections}
\label{Section:WeylConnections}

Given a spacetime $(\tilde{\mathcal{M}},\tilde{\bmg})$, a \emph{Weyl
  connection} $\hat{\bmnabla}$ is a torsion-free connection
satisfying
\[
\hat{\nabla}_a \tilde{g}_{bc} = -2 f_a \tilde{g}_{bc}
\]
for some smooth 1-form $\bmf$. Conformal geodesics allow to single out
a \emph{canonical} Weyl connection. More precisely, given a solution
$(\bmx(\tau),\bmb(\tau))$ to the conformal geodesic equations
\eqref{ConformalGeodesicEquation1}-\eqref{ConformalGeodesicEquation2}
one can set $\bmf=\tilde{\bmb} - \Theta^{-1} \mathbf{d} \Theta$ so that using the transformation rules 
of the conformal geodesic equations under changes of connection one concludes
that
\begin{equation}
\hat{\nabla}_{\dot{\bmx}} \dot{\bmx} =0, \qquad \hat{\bmL}(\dot{\bmx},
  \cdot) =0, \qquad \hat{\nabla}_{\dot{\bmx}} \bme_\bma =0,
\label{WeylConnectionGaugeConditions}
\end{equation}
where $\hat{\bmL}$ denotes the Schouten tensor of the Weyl connection
$\hat{\bmnabla}$. Hence, the curve $\bmx(\tau)$ is a (standard) geodesic with
respect to $\hat{\bmnabla}$ and the frame $\{ \bme_\bma \}$ is
parallely propagated with respect to $\hat{\bmnabla}$ along the curve. For more details on Weyl
connections and their relation to conformal geodesics, see
\cite{Fri95,Fri03c}.

\section{Time symmetric hypersurfaces of the extremal
  Reissner-Nordstr\"om spacetime}
\label{Section:TimeSymmetricSlices}

As a consequence of the timelike nature of the static Killing vector
$\bmpartial_t$ everywhere (except at the horizon), it
follows that the maximal analytical extension of the spacetime admits
two types of \emph{time symmetric hypersurfaces} ---that is, hypersurfaces
with vanishing extrinsic curvature: 
\begin{enumerate}[(a)]
\item The time symmetric hypersurfaces $\tilde{\mathcal{S}}$ contained in the regions $\widetilde{I}$.
These hypersurfaces are \emph{asymptotically Euclidean} at $i^0$ and have a
\emph{cylindrical asymptotic end} $c^0$. The cylindrical asymptotic end is
shared by other black hole solutions, most notably the extremal Kerr
solution ---see e.g. \cite{DaiGab11} for further details.
\item The time symmetric hypersurfaces contained in the regions $\widetilde{III}$, to be denoted by
$\tilde{\regionIIIslices}$, which start at $i^+$ and end at the
singularity. The conformal diagram of the extremal
Reissner-Nordstr\"om spacetime suggests that these hypersurfaces are
some sort of \emph{degenerate} hyperboloid. 
\end{enumerate}
The location of these hypersurfaces is identified in Figure
\ref{ConformalDiagrameRN}, left. 

\medskip
By contrast, the negative mass extremal Reissner-Nordstr\"om contains
a single asymptotically Euclidean time symmetric hypersurface $\tilde{\mathcal{S}}'$ which,
in many senses, resembles the case (a) above ---see also Figure
\ref{ConformalDiagrameRN}, right.

\subsection{The conformal constraints}
\label{Section:ConformalConstraints}

In what follows we analyse the properties of the hypersurfaces
$\tilde{S}$ and $\tilde{\regionIIIslices}$ in some detail from the
perspective of the conformal Einstein field equations. The conformal
Einstein-Maxwell field equations have been discussed in
\cite{Fri91,LueVal12,Sim92}. In particular, the constraint equations
implied by the conformal Einstein-Maxwell equation in a spacelike
hypersurface have been discussed in \cite{Sim92}. In view of the
applications of this article, attention will be restricted to the
\emph{time symmetric conformal constraint equations}.

\medskip
In the remaining of this section, let $\mathcal{S}$ denote an arbitrary
hypersurface of an unphysical spacetime $(\mathcal{M},\bmg)$ and let
$\Xi$ denote the associated conformal factor ---i.e. $\bmg =
\Xi^2\tilde{\bmg}$ where $\tilde{\bmg}$ denotes, as usual, the
physical spacetime metric. 
Recall that Maxwell's equations are conformally invariant hence
the \emph{unphysical Faraday tensor} $F_{ab}$ is given by
\[
F_{ab} = \tilde{F}_{ab}.
\]

The restriction of $\Xi$ to $\mathcal{S}$
will be denoted by $\omega$. Let $\bmn$ indicate the $\bmg$-unit normal to
$\mathcal{S}$, and let $\{ \bme_\bma\}$ denote a $\bmg$-orthonormal
frame such that $\bme_0 = \bmn$. Consequently, $\{\bme_\bmi\}$ constitutes a
basis of orthonormal spatial vectors intrinsic to $\mathcal{S}$. If
$\bmh$ denotes the intrinsic metric of $\mathcal{S}$ implied by
$\bmg$, one has that
\[
h_{\bmi\bmj}\equiv \bmh(\bme_\bmi,\bme_\bmj) = -\delta_{\bmi\bmj}.
\]

In addition to $\omega$ as defined in the previous paragraph, the
conformal constraint equations are expressed in terms of the fields
\begin{eqnarray*}
s  && \equiv \tfrac{1}{4} \nabla^a \nabla_a \Xi + \tfrac{1}{4}L \Xi,\\
L_{\bmi\bmj} && \mbox{(the spatial part of the Schouten tensor
  of $\bmg$)},\\ 
d_{\bmi\bmj} && = \Xi^{-1} C_{0\bmi 0 \bmj} \phantom{X} \mbox{(the electric part of the rescaled Weyl tensor)}, \\
l_{\bmi\bmj}  && =r_{\bmi\bmj} - \tfrac{1}{4} r h_{\bmi\bmj}
\phantom{X} \mbox{(the Schouten tensor of $\bmh$)},\\
E_\bmi && = F_{0\bmi} \mbox{(the electric part of the Faraday tensor)}, \\
\mu  &&  \equiv E_\bmi E^\bmi \phantom{X}\mbox{ (the unphysical energy density of
  the Faraday tensor)},
\end{eqnarray*}
where $r_{\bmi\bmj}$ denotes the components of the Ricci tensor of the
3-metric $\bmh$ and $r$ is its Ricci scalar.

\medskip
For hypersurfaces which are  time symmetric and maximal in both the
physical and unphysical spacetimes, the conformal
Einstein constraint equations imply 
\begin{subequations}
\begin{eqnarray}
&& D_\bmi D_\bmj \omega = -\omega L_{\bmi\bmj} + s
h_{\bmi\bmj} + \omega^3 E_{\{\bmi} E_{\bmj\}}, \label{ConformalConstraint1} \\
&& 6 \omega s - 3 D_\bmk \omega D^\bmk
 \omega=0, \label{ConformalConstraint8}\\
&& l_{\bmi\bmj} = \omega d_{\bmi\bmj} + L_{\bmi\bmj}, \label{ConformalConstraintExtra}
\end{eqnarray}
\end{subequations}
where $E_{\{\bmi} E_{\bmj\}}$ denotes the $\bmh$-tracefree part of $E_\bmi E_\bmj$.
The above equations imply, in particular, the \emph{time symmetric Hamiltonian constraint}
\[
2\omega D_\bmi D^\bmi \omega - 3 D_\bmi \omega D^\bmi\omega
+ \tfrac{1}{2} \omega^2 r  = \omega^4 \mu.
\]
The electric field $E_\bmi$ satisfies the constraint
\begin{equation}
D^\bmi E_\bmi =0,
\label{MaxwellConstraint}
\end{equation}
where 
\[
E_\bmi = \omega^{-1} \tilde{E}_\bmi.
\]

The constraint equations
\eqref{ConformalConstraint1}-\eqref{ConformalConstraintExtra} can be
solved to yield $d_{\bmi\bmj}$ and $L_{\bmi\bmj}$ in terms of $\omega$
and its derivatives, intrinsic geometric fields and the electric
field. One has that:
\begin{subequations}
\begin{eqnarray}
&& d_{\bmi\bmj} = \frac{1}{\omega^2} D_{\{\bmi} D_{\bmj\}} \omega +
\frac{1}{\omega}l_{\{\bmi\bmj\}} + \omega E_{\{\bmi} E_{\bmj\}}, \label{CFEData1}\\
&& L_{\bmi\bmj} = -\frac{1}{\omega} D_{\{\bmi} D_{\bmj\}} \omega
+\tfrac{1}{12} r h_{\bmi\bmj} - \omega^2 E_{\{\bmi} E_{\bmj\}} \label{CFEData2}
\end{eqnarray}
\end{subequations}
These equations will be key for the remaining analysis of this section.

\subsection{Basic conformal extensions of the time symmetric initial
  hypersurfaces}
\label{Section:BasicConformalExtensions}

A direct computation shows that the intrinsic metric $\tilde{\bmh}$
implied by the extremal
Reissner-Nordstr\"om metric \eqref{eRNIsotropic} on any time symmetric
hypersurface of the spacetime is formally given by 
\begin{equation}
\tilde{\bmh} = -\left(1+\frac{m}{r} \right)^2 (\mathbf{d} r \otimes
\mathbf{d} r + r^2 \bmsigma),
\label{3MetricRegionI}
\end{equation}
 The key observation from expression \eqref{3MetricRegionI} is that the
3-metrics of the time symmetric hypersurfaces are conformally
flat. This property simplifies many of the computations in the
sequel. One also finds that the initial electric field is given by 
\begin{equation}
\tilde{\bmE} = \pm \frac{m}{r(r+m)}\mathbf{d}r .
\end{equation}

\medskip
The precise properties of the initial data  depend on the signs of $m$ and $r$. 

\subsubsection{The time symmetric hypersurface of region $\widetilde{I}$}
The intrinsic metric of the time symmetric hypersurface
$\tilde{\mathcal{S}}$ is given by the metric $\tilde{\bmh}$ of \eqref{3MetricRegionI} with the
conditions $m>0$ and $r>0$. The asymptotically Euclidean end of $\tilde{\mathcal{S}}$ corresponds
to the condition $r\rightarrow \infty$. Thus, it is natural to introduce
the coordinate 
\[
\rho\equiv 1/r.
\] 
A computation shows that
\begin{eqnarray*}
&& \bmh \equiv \omega^2 \tilde{\bmh} = - (\mathbf{d}\rho\otimes \mathbf{d}\rho + \rho^2 \bmsigma) =-\bmdelta,
\end{eqnarray*}
with 
\begin{equation}
\omega = \frac{\rho^2}{(1+m\rho)}.
\label{omegaSi0}
\end{equation}

The rescaled (unphysical) electric field is given by
\[
\bmE = \omega^{-1} \tilde{\bmE} = \mp \frac{m}{\rho^2}\mathbf{d}\rho . 
\]
The metric $\bmh=-\bmdelta$ is clearly regular at $\rho=0$. Thus, one
obtains a conformal extension $(\mathcal{S}_{i^0},-\bmdelta)$ of the hypersurface
$\widetilde{\mathcal{S}}$ for which $i^0\equiv \{ p\in
\mathcal{S}_{i^0}\,|\,\rho=0\}$ corresponds to the point at infinity. One
can readily verify that
\[
\omega|_{i^0}=0, \qquad \mathbf{d}\omega|_{i^0} =0, \qquad
\mbox{\textbf{Hess}}\, \omega|_{i^0} = 2 \bmdelta.
\]
In order to analyse the behaviour of the fields $d_{\bmi\bmj}$ and
$L_{\bmi\bmj}$, as given by equations \eqref{CFEData1}-\eqref{CFEData2},  in a neighbourhood of $i^0$ it is convenient
to consider on $\mathcal{S}_{i^0}$ a system of normal Cartesian coordinates $(x^\bmi)$ centred
at $i^0$ ---that is, $x^\bmi(i^0)=0$--- with $\rho^2 =
\delta_{\bmi\bmj} x^\bmi x^\bmj$. Moreover, on $\mathcal{S}_{i^0}$
consider a frame $\bme_\bmi$ such that $\bme_\bmi =
\bmpartial/\bmpartial x^\bmi$. By construction one readily has that
$\bmh(\bme_\bmi,\bme_\bmj) =-\delta_{\bmi\bmj}$. Using equations \eqref{CFEData1}-\eqref{CFEData2}, keeping in mind
that $\bmh$ is flat so that $r_{\bmi\bmj}=r=0$, it follows that
\begin{subequations}
\begin{eqnarray}
&& d_{\bmi \bmj} = - \frac{3 m }{\rho^5 (1+m\rho)}
x_{\{\bmi}x_{\bmj\}}, \label{SingularitySDetailed1}\\
&& L_{\bmi\bmj} = \frac{3m}{\rho^3 (1+m\rho)^2} x_{\{\bmi}x_{\bmj\}},  \label{SingularitySDetailed2}\\
&& E_\bmi = \pm \frac{m}{\rho^3} x_\bmi,  \label{SingularitySDetailed3}
\end{eqnarray}
\end{subequations}
with $x_\bmi \equiv -\delta_{\bmi\bmj} x^\bmj$. Consequently, one has that
\begin{equation}
d_{\bmi\bmj} = O(\rho^{-3}), \qquad L_{\bmi\bmj} =
O(\rho^{-1}), \qquad E_\bmi = O(\rho^{-2}) \qquad \mbox{on} \;\; \mathcal{S}_{i^0}.
\label{SingularityS}
\end{equation}
One sees that the data is singular at $i^0$. This singular behaviour
is well known in the case of vacuum spacetimes ---see
e.g. \cite{Fri88,Fri98a}. 

\medskip
Now, the cylinder-like asymptotic end of the hypersurface
$\tilde{\mathcal{S}}$ corresponds to the condition $r\rightarrow 0$. A direct computation shows that the set
of points for which $r=0$ lies at infinity with respect to the metric
$\tilde{\bmh}$. Hence, in this case it is natural to use the coordinate $r$ to
construct a conformal extension of $\tilde{\mathcal{S}}$ near the
cylinder end by letting
\[
\bmh = \varpi^2 \tilde{\bmh} =  - (\mathbf{d} r\otimes \mathbf{d} r + r^2 \bmsigma) =-\bmdelta,
\]
with
\[
\varpi = \frac{r}{r+m}.
\]
Clearly $\bmh$ is smooth at $r=0$, so that one obtains a smooth
conformal extension $(\mathcal{S}_{c^0},-\bmdelta)$ of
$(\tilde{\mathcal{S}},\tilde{\bmh})$.  Letting $c^0\equiv \{ p\in
\mathcal{S}_{c^0} \,|\,  r=0 \}$ one readily sees that
\[
\varpi|_{c^0}=0, \qquad \mathbf{d}\varpi|_{c^0}\neq 0.
\]
Thus, the behaviour at the cylinder end resembles that of an hyperboloid. As in the
analysis of $i^0$ it is convenient to introduce a Cartesian system of
normal coordinates $(x^\bmi)$ centred at $c^0$ and an associated
orthonormal frame $\{ \bme_\bmi\}$ with $\bme_\bmi =
\bmpartial_\bmi$. As a consequence of the flatness of the conformal metric, it is
easy to compute the fields $d_{\bmi\bmj}$ and $L_{\bmi\bmj}$ on
$\mathcal{S}_{c^0}$. Using expressions \eqref{CFEData1}-\eqref{CFEData2} one
readily finds that 
\begin{eqnarray*}
&& d_{\bmi\bmj} = - \frac{3m}{r^4(r+m)} x_{\{\bmi} x_{\bmj\}}, \\
&& L_{\bmi\bmj} = \frac{3m}{r^3(r+m)^2} x_{\{\bmi} x_{\bmj\}}, \\
&& E_\bmi = \pm \frac{m}{r^3} x_\bmi.
\end{eqnarray*}
The above data is singular at $i^+$. More precisely, one has that
\begin{equation}
d_{\bmi\bmj} = O(r^{-2}), \qquad L_{\bmi\bmj} = O(r^{-1}), \qquad
E_\bmi = O(r^{-2}) \qquad \mbox{on} \phantom{X} \mathcal{S}_{c^0}. 
\label{SingularityCylinder}
\end{equation}

\subsubsection{The time symmetric slice of region
  $\widetilde{I}'$}
The intrinsic metric of the time symmetric hypersurface
$\tilde{\mathcal{S}}'$ is given by the metric $\tilde{\bmh}$ of \eqref{3MetricRegionI} with the
conditions $m<0$ and $r>|m|$.  This hypersurface
has an asymptotically Euclidean end corresponding to the condition
$r\rightarrow \infty$. A conformal extension can be obtained in a
similar way to what was done for the asymptotically Euclidean end of
$\tilde{\mathcal{S}}$ by introducing the coordinate $\rho=1/r$. In particular, the conformal factor
\[
\omega = \frac{\rho^2}{(1+m\rho)}.
\]
is formally identical to that of $\omega$ as given in
\eqref{omegaSi0}. The corresponding conformal extension including
$\rho=0$ will be denoted by $(\mathcal{S}'_{i^0},-\bmdelta)$. It
follows that one obtains the singular behaviour
\begin{equation}
d_{\bmi\bmj} = O(\rho^{-3}), \qquad L_{\bmi\bmj} =
O(\rho^{-1}), \qquad E_\bmi = O(\rho^{-2}) \qquad \mbox{on} \;\; \mathcal{S}'_{i^0}.
\label{SingularitySbar}
\end{equation}

\subsubsection{The time symmetric slice of region
  $\widetilde{III}$}
Finally, we consider the case of the time symmetric hypersurface
$\tilde{\regionIIIslices}$ in region $\widetilde{III}$. In this case the intrinsic
metric is given by the expression for $\tilde{\bmh}$ of equation
\eqref{3MetricRegionI} with the condition $r<0$. This situation is
completely analogous to the discussion of the cylinder-like end of
region $\widetilde{I}$. In particular, one can use the coordinate $r$
to construct a conformal extension $(\regionIIIslices,-\bmdelta)$ with a
conformal factor 
\[
\varpi = \frac{r}{r+m},
\]
from where one concludes the same singular behaviour as in
\eqref{SingularityCylinder} at $i^+ =\{ p \in \regionIIIslices \,|\, r=0 \}
$.

\subsection{The conformal isometry between the time symmetric
  hypersurfaces}
\label{Section:ConformalIsometryHypersurfaces}

In order to gain further intuition into the behaviour of the extremal
Reissner-Nordstr\"om spacetime around $c^0$ and $i^+$, it is convenient to analyse the effects of the
conformal isometry $\iota$ on the various time symmetric hypersurfaces.

\subsubsection{The hypersurface $\tilde{\mathcal{S}}$}
Taking into account that $\rho\equiv 1/r$ and that $\varrho\equiv
m^2/r$, it follows readily that $\varrho=m^2\rho$ and that
\[
 m^4 (\mathbf{d} \rho \otimes \mathbf{d} \rho + \rho^2 \bmsigma ) =
\mathbf{d}\varrho \otimes \mathbf{d}\varrho + \varrho^2 \bmsigma.
\]
Hence,
\[
\iota_* \bmh = \frac{1}{m^4}(\mathbf{d} r \otimes \mathbf{d} r + r^2 \bmsigma).
\]
Thus, under the conformal isometry the limit $\rho\rightarrow 0$, respectively $\varrho\rightarrow 0$, correspond to
$r\rightarrow 0$. Thus one has that
\[
\iota(\mathcal{S}_{i^0}) = \mathcal{S}_{c^0},
\]
and, in particular
\[
\iota(i^0) = c^0.
\]

\subsubsection{The hypersurfaces $\regionIIIslices$ and
  $\widetilde{\mathcal{S}}'$}

Using the expressions of the previous paragraph, and recalling that the
conformal isometry $\iota$ maps $\widetilde{I}'$ into
$\widetilde{III}\cup \mathscr{H}^+$, one concludes that
\[
\iota(\mathcal{S}')= \regionIIIslices,
\]
and, in particular
\[
\iota(i^{\prime 0}) = i^+. 
\]

\subsection{Some remarks}
The singular behaviour of the conformal fields $d_{\bmi\bmj}$ and
$L_{\bmi\bmj}$ at $i^0$ and $i^{\prime 0}$ given, respectively, by
\eqref{SingularityS} and \eqref{SingularitySbar} is the main technical difficulty in the analysis
of the conformal field equations in this region of spacetime ---this is
sometimes known as the \emph{problem of spatial infinity}. 
For vacuum spacetimes it has been shown in \cite{Fri98a}
how one can introduce a conformal representation of this
region of spacetime in which the the equations and
their data are regular ---the so-called \emph{cylinder at spatial
  infinity}, see the discussion in the introduction. We will show
 in Section \ref{Section:CylindersHorizon} that this construction can be extended, with
minor modifications, to the electrovacuum case. 

The slightly milder
singular behaviour at $c^0$ and $i^+$ observed in \eqref{SingularityCylinder} suggests
that a similar regular conformal representation could be introduced
for this part of the (positive mass) extremal Reissner-Nordstr\"om
spacetime. Moreover, the correspondence between the asymptotically
Euclidean ends and the \emph{singular hyperboloidal ends} $c^0$ and
$i^+$ discussed in Section
\ref{Section:ConformalIsometryHypersurfaces} raises the following expectation:
The regular representations can be simply mapped
from the cylinders at infinity of $i^0$ and $i^{\prime 0}$ rather than
built from scratch.

\section{The cylinder at spatial infinity for the extremal
  Reissner-Nordstr\"om spacetime}
\label{Section:CylinderSpatialInfinity}

The purpose of this section is to provide a discussion of the
construction of the cylinder at infinity for the asymptotically
Euclidean ends of the time symmetric hypersurfaces $\tilde{\mathcal{S}}$ and
$\tilde{\mathcal{S}}'$ of the extremal Reissner-Nordstr\"om
spacetime.  This
construction follows closely the one for the Schwarzschild spacetime in
\cite{Fri98a}. The construction is local to a neighbourhood of spatial
infinity, and thus, independent of the sign of the mass
parameter. Accordingly, for the ease of presentation, we consider the
positive and negative mass cases simultaneously. 

\subsection{The bundle space $\mathcal{C}_{a}$}
The first step in the construction of the cylinder at infinity
consists of the \emph{blowing up} of spatial infinity $i^0$ to the
2-sphere $\mathbb{S}^2$. Technically, the implementation of this idea
requires the introduction of a bundle space $\mathcal{C}_{a}$. The
bundle space $\mathcal{C}_{a}$ and its extensions play a central role
in our subsequent analysis. Hence we give a brief overview of the
construction of $\mathcal{C}_{a}$ and these extensions below ---for
details the reader is referred to \cite{Fri98a}.

Let $\mathcal{B}_a(i^0)\subset \mathcal{S}_{i^0}$ be the open ball of
radius $a$ centred at $i^0$ for some sufficiently small $a>0$. Further
let $SU(\mathcal{B}_a(i^0))$ is the bundle of normalised spin frames
over $\mathcal{B}_a(i^0)$ with structure group $SU(2,\mathbb{C})$. To
obtain the manifold $\mathcal{C}_a$ one starts by choosing a fixed
normalised spin dyad $\delta^*_\bmA$ at $i^0$. Any other spin frame is
of the form $\delta_\bmA(t)=\delta^*_\bmB t^\bmB{}_\bmA$ with
$t=(t^\bmB{}_\bmA) \in SU(2,\mathbb{C})$. For a given value of $t$,
the spin frame $\delta_\bmA(t)$ gives rise to an orthonormal frame
\[ \bme_\bmi(t) \equiv \sigma_\bmi{}^{\bmA\bmB} \delta_\bmA(t)
\delta_\bmB(t),
\] where $\sigma_\bmi{}^{\bmA\bmB}$ indicates the \emph{spatial
Infeld-van der Waerden symbols}. For some values of $t$, the frame
vector $\bme_3(t)= \sigma_3{}^{\bmA\bmB} \delta_\bmA(t)
\delta_\bmB(t)$ corresponds to the radial vector at $i^0$. Keeping $t$
fixed, one then constructs on $\mathcal{B}_a(i^0)$ the $\bmh$-geodesic starting at $i^0$ with initial
tangent vector $\bme_3(t)$. As a consequence of the flatness of the
conformal metric $\bmh$, the coordinate $\rho$ is an affine parameter
of this geodesic that vanishes at $i^0$. The spin dyad $\delta_\bmA$
is then propagated along the geodesic. For a particular value of
$\rho$ the spin dyad so constructed will be denoted by
$\delta_\bmA(\rho,t)$.  One then sets
\begin{equation}
\mathcal{C}_a \equiv \left\{ \delta_\bmA(\rho,t)\in SU(\mathcal{B}_a(i^0)) \, |\, 0\leq \rho <a \right\}.
\label{DefinitionCa}
\end{equation}
The boundary of the bundle manifold $\mathcal{C}_a$ consists of the
set 
\[
\cylinderinfinity^0\equiv \{ \delta_\bmA(\rho,t) \in \mathcal{C}_a \; | \;
\rho=0\}.
\]
It can be verified that $\cylinderinfinity^0\simeq SU(2,\mathbb{C})$, so that the components of
the boundary can be regarded as the blow up of the point at infinity
$i^0$. 

\medskip
For more details on the various aspects of this construction, the
reader is referred to \cite{Fri98a}, Section 3. An alternative,
abridged discussion is given in \cite{AceVal11}.

\subsubsection{Lifts to $\mathcal{C}_a$}
\label{Section:LiftsBundleSpace} 

Any smooth spinor field on $\mathcal{B}_a(i^0)\subset \mathcal{S}_{i^0}$ is represented on
$\mathcal{C}_a$ by a \emph{spinor valued function} given
at $\delta_\bmA \in \mathcal{C}_a$ by the components of the
spinor in the dyad defined by $\delta_\bmA$. This procedure will be
referred to as the \emph{lift} of the spinor field. The lift to
$\mathcal{C}_a$ of any symmetric valence 2 spinor on
$\mathcal{B}_a(i)$ can be spanned in terms of symmetric spinors $x_{\bmA\bmB}$,
$y_{\bmA\bmB}$, $z_{\bmA\bmB}$ such that
\[
x_\bmA{}^\bmQ x_{\bmB\bmQ}=\tfrac{1}{2} \epsilon_{\bmA\bmB}, \quad x_{\bmB\bmQ} y_\bmA{}^\bmQ =
  \tfrac{1}{\sqrt{2}} y_{\bmA\bmB}, \quad x_{\bmB\bmQ} z_\bmA{}^\bmQ = -
  \tfrac{1}{\sqrt{2}} z_{\bmA\bmB}, \quad y_\bmA{}^\bmQ z_{\bmB\bmQ} =
  -\tfrac{1}{2\sqrt{2}}x_{\bmA\bmB} + \tfrac{1}{4}\epsilon_{\bmA\bmB},
\]
and $y_\bmA{}^\bmQ y_{\bmB\bmQ}= z_\bmA{}^\bmQ z_{\bmB\bmQ}=0$. Higher valence spinors can be
spanned by suitable combinations of these spinors and the totally
antisymmetric spinor $\epsilon_{\bmA\bmB}$ ---\cite{Fri98a,FriKan00}.

\subsubsection{Vector fields on $\mathcal{C}_a$}
\label{Section:VectorFieldsCa}

The manifold $\mathcal{C}_a$ has a dimension more than
$\mathcal{S}_{i^0}$. This extra dimension corresponds to the action of the
subgroup $U(1)$ of $SU(2,\mathbb{C})$. In what follows, we will use
$t^\bmA{}_\bmB \in SU(2,\mathbb{C})$ and $\rho$ as coordinates on
$\mathcal{C}_a$. In order to be able to compute derivatives on $\mathcal{C}_a$, one considers a basis $\{\bmX_+,\,
\bmX_-,\, \bmX\}$ of the Lie algebra
$\mathfrak{su}(2,\mathbb{C})$, such that $\bmX$ is the generator of
$U(1)$ and one has the commutation relations
\[
[\bmX,\bmX_+]=2\bmX_+, \quad [\bmX,\bmX_-]=-2\bmX_-, \quad [\bmX_+,\bmX_-]=-\bmX,
\]
with $\bmX_+$ and $\bmX_-$ complex conjugates of each
other. These vector fields are extended to $\mathcal{C}_a$
by the requirements
\[
[\bmpartial_\rho, \bmX]=0, \quad [\bmpartial_\rho, \bmX_+]=0, \quad [\bmpartial_\rho, \bmX_-]=0.
\]
The vector fields $\{\bmpartial_\rho,\, \bmX, \, \bmX_+,\, \bmX_-\}$ constitute
a frame field on $\mathcal{C}_a$. A function $f$ is said to
have \emph{spin weight} $s$ if ${\bm X} f = 2sf$, with $s$ an integer. Any
spinor-valued function on $\mathcal{C}_a$ has a well defined
spin weight. To complete the discussion, one
requires to consider forms $\bmalpha^+$, $\bmalpha^-$ and
$\bmalpha$ which annihilate the the vector field $\partial_\rho$ and,
in addition, satisfy
\[
\langle \bmalpha^+  , \bmX_+ \rangle = \langle \bmalpha^-
, \bmX_- \rangle= \langle \bmalpha , \bmX \rangle=1.
\]
The normalisation conventions being used are such that $2(\bmalpha^+
\otimes \bmalpha^- + \bmalpha^- \otimes \bmalpha^+)$ pulls back to
$\bmsigma$,  the
standard metric on $\mathbb{S}^2$. 

\subsubsection{Frame fields, solder forms and connection forms}
\label{Section:SolderForms}

The vector fields and 1-forms on $\mathcal{C}_a$ introduced
in the previous subsection will be used to span the following frame fields and
corresponding solder forms:
\begin{eqnarray*}
&& {\bm e}_{\bmA\bmB} = x_{\bmA\bmB} \partial_\rho + \frac{1}{\rho} z_{\bmA\bmB}{\bm
  X}_+ + \frac{1}{\rho} y_{\bmA\bmB} {\bm X}_-,  \\
&& {\bm \sigma}^{\bmA\bmB} = - x^{\bmA\bmB} \mbox{d}\rho -2 \rho y^{\bmA\bmB} {\bm \alpha}^+
-2\rho z^{\bmA\bmB} {\bm \alpha}^-. 
\end{eqnarray*}
The above fields have been chosen so that 
\[
\bmh = h_{\bmA\bmB\bmC\bmD} \sigma^{\bmA\bmB} \otimes \sigma^{\bmC\bmD}, \qquad \langle {\bm
  \sigma}^{\bmA\bmB}, {\bm e}_{\bmC\bmD} \rangle = h^{\bmA\bmB}{}_{\bmC\bmD},  
\]
where in a slight abuse of notation $\bmh$ denotes the lift to
$\mathcal{C}_a$ of the conformal metric $\bmh=-\bmdelta$, and
$h_{\bmA\bmB\bmC\bmD}\equiv -\epsilon_{\bmA(\bmC} \epsilon_{\bmD)\bmB}$ is the spinorial
counterpart of $-\delta_{\bmi\bmj}$. The associated spin connection coefficients
$\gamma_{\bmA\bmB\bmC\bmD}$ can be computed using the spinorial version of the
Cartan structure equations. One has that
\[
\gamma_{\bmA\bmB\bmC\bmD} = \frac{1}{2\rho} (\epsilon_{\bmA\bmC} x_{\bmB\bmD} +
\epsilon_{\bmB\bmD} x_{\bmA\bmC}).
\]

\medskip
Covariant differentiation on
$\mathcal{C}_a$ is performed using the standard rules. Let
$F$ denote the lift to $\mathcal{C}_a$ of a smooth
function, $f$, on $\mathcal{B}_a(i^0)$. The covariant derivative $D_{AB}f$
is represented on $\mathcal{C}_a$ by $\bme_{\bmA\bmB}
F$. In order to ease the notation, in what follows the same symbol
will be used to denote a function on $\mathcal{B}_a(i^0)$ and its lift to
$\mathcal{C}_a$. Using this convention, let $\mu_{\bmA\bmB}$ denote the lift to
$\mathcal{C}_a$ of the spinorial field $\mu_{AB}$ on
$\mathcal{B}_a(i^0)$.  The lift of the covariant derivative $D_{AB}\mu_{CD}$
is then given by
\[
D_{\bmA\bmB} \mu_{\bmC\bmD} = \bme_{\bmA\bmB} \mu_{\bmC\bmD} - \gamma_{\bmA\bmB}{}^\bmP{}_\bmC
\mu_{\bmP\bmD} - \gamma_{\bmA\bmB}{}^\bmP{}_\bmD\mu_{\bmC\bmP}.
\]
Similar expressions hold for higher valence spinors.

\subsection{The extended bundle space $\mathcal{C}_{a,\kappa}$}
\label{Section:ExtendedBundleSpace}

The second step in the construction of the cylinder at spatial
infinity consists of the introduction of an extended bundle space in
which the spin dyads $\{ \delta_\bmA \}$ are rescaled by a certain
factor so that the components of singular fields at $i^0$ become
regular. Given a non-negative smooth function $\kappa$ one defines the
\emph{extended bundle space}
\[
\mathcal{C}_{a,\kappa} \equiv \{ \kappa^{1/2} \delta_\bmA \; | \; \delta_\bmA
\in \mathcal{C}_a \}. 
\]

\medskip
In what follows, let $\phi_{ABCD}$, $\Theta_{ABCD}$ and $\varphi_{AB}$
denote the spinorial counterparts of the fields $d_{\bmi\bmj}$,
$\hat{L}_{\bmi\bmj}$ and $E_\bmi$ as discussed in Section
\ref{Section:ConformalConstraints}. Following the conventions of the
previous sections, we denote their lifts to the bundle space
$\mathcal{C}_a$ by $\phi_{\bmA\bmB\bmC\bmD}$,
$\Theta_{\bmA\bmB\bmC\bmD}$ and $\varphi_{\bmA\bmB}$.  Under the
rescaling $\delta_\bmA\mapsto \kappa^{1/2} \delta_\bmA$, the latter
fields can be seen to transform as 
\[
\phi_{\bmA\bmB\bmC\bmD} \mapsto \kappa^3 \phi_{\bmA\bmB\bmC\bmD},
\qquad 
\Theta_{\bmA\bmB\bmC\bmD} \mapsto \kappa^2 \Theta_{\bmA\bmB\bmC\bmD},
\qquad
\varphi_{\bmA\bmB} \mapsto \kappa^2 \varphi_{\bmA\bmB} 
\]
---see \cite{Fri98a} for further details. In order to choose $\kappa$
appropriately one needs to consider the lift of the above fields to
$\mathcal{C}_a$. A calculation using expressions \eqref{SingularitySDetailed1}-\eqref{SingularitySDetailed3} shows that 
\[
\phi_{\bmA\bmB\bmC\bmD} = -
\frac{\displaystyle 6m}{\rho^3(\displaystyle
  1+m\rho)}\epsilon^2_{\bmA\bmB\bmC\bmD}, 
\qquad  \Theta_{\bmA\bmB\bmC\bmD} = \frac{\displaystyle 6m}{\displaystyle
  \rho^2 (1+m\rho)^2}\epsilon^2_{\bmA\bmB\bmC\bmD}, 
\qquad \varphi_{\bmA\bmB} = \mp \frac{m}{\rho^2} x_{\bmA\bmB},
\]
where $\epsilon^2_{\bmA\bmB\bmC\bmD} \equiv \tfrac{1}{2} x_{(\bmA\bmB}
x_{\bmC\bmD)}$. Thus, in order to obtain rescaled fields which are
finite at $\cylinderinfinity^0$, it follows that one should chose
$\kappa=O(\rho)$. For simplicity, we make the choice
\begin{equation}
\kappa =\rho.
\label{Choicekappa}
\end{equation}

\medskip
In the sequel, we will also require the transformation behaviour of
the frame fields $\bme_{\bmA\bmB}$, the soldering forms
$\bmsigma^{\bmA\bmB}$ and the spatial connection coefficients $\gamma_{\bmA\bmB\bmC\bmD}$. These are given by
\[
\bme_{\bmA\bmB} \mapsto \kappa \bme_{\bmA\bmB}, \qquad
\bmsigma^{\bmA\bmB} \mapsto \kappa^{-1} \bmsigma^{\bmA\bmB}, \qquad
\gamma_{\bmA\bmB\bmC\bmD} \mapsto \kappa \gamma_{\bmA\bmB\bmC\bmD}
-\tfrac{1}{2}(\epsilon_{\bmA\bmC} \bme_{\bmB\bmD}\kappa +
\epsilon_{\bmB\bmD} \bme_{\bmA\bmC} \kappa).
\]
Thus, in particular, one has that the lift of the 3-metric $\bmh=-\bmdelta$ satisfies 
\[
\bmh \mapsto \frac{1}{\rho^2} \mathbf{d}\rho \otimes \mathbf{d}\rho +
2(\bmalpha^+ \otimes \bmalpha^- + \bmalpha^- \otimes \bmalpha^+).
\]
This a feature not often appreciated of the construction of the
cylinder at spatial infinity. Namely, that it renders a 3-metric which
is singular at $\rho=0$ ---or in other words, the cylinder at spatial
infinity is at an infinite distance as measured by the 3-metric
$\kappa^{-1}\bmh$ so that the representation is not metrically compact. This does not cause problems with the
conformal field equations as the metric (and the soldering forms) do
not appear as unknowns in the equations ---see the discussion in
Section \ref{Section:CFE}.

\subsection{Spacetime gauge considerations}
\label{Section:SpacetimeGaugeConsiderations}
The third step in the construction of the cylinder at spatial
infinity consists of the implementation of a suitable spacetime gauge to analyse
the evolution of initial data prescribed on $\mathcal{C}_a$. Again,
following the ideas of \cite{Fri98a}, we rely on a gauge based
on a congruence of conformal geodesics. 

\subsubsection{Initial data for the congruence of conformal geodesics}
Initial data for the congruence of conformal geodesics is prescribed
for a given  $p\in \mathcal{B}_a(i^0)\setminus \{ i^0 \}$ so that
\begin{equation}
\bmx_\star = \bmx(p), \qquad \dot{\bmx}_\star \perp \tilde{\mathcal{S}}_{i^0}, \qquad
\tilde{\bmb}_\star = \omega^{-1} \mathbf{d} \omega, \qquad \langle
\tilde{\bmb}, \dot{\bmx}
\rangle_\star =0.
\label{ConformalGeodesicsData}
\end{equation}
The above initial data is supplemented by the following
choice of data for the conformal factor ---cf. equations
\eqref{ThetaEvolutionData1}-\eqref{ThetaEvolutionData2}:
\[
\Theta_\star = \kappa^{-1} \omega, \qquad \dot{\Theta}_\star =0, \qquad
\ddot{\Theta}_\star =
-\frac{\kappa}{2\omega}\bmdelta(\mathbf{d}\omega,\mathbf{d}\omega) +
\kappa^{-3} \omega^3 \mu_\star , 
\]
with $\kappa$ given as in \eqref{Choicekappa} and $\mu_\star$ the
initial value of the energy density of the Maxwell field.

\subsubsection{Gauge conditions associated to the Weyl connection}
\label{Section:WeylGaugeConditions}

As it will be seen in the next section, we will consider conformal
field equations for the Einstein-Maxwell system which are expressed in
terms of the Weyl connection $\hat{\bmnabla}$ associated to the congruence of conformal
geodesics with initial data given by
\eqref{ConformalGeodesicsData}. In addition to this connection, we
also need to keep track of the Levi-Civita connections
$\tilde{\bmnabla}$ and $\bmnabla$ of the metrics $\tilde{\bmg}$ and
$\bmg= \Theta^2 \tilde{\bmg}$, respectively. These three connections are related to each
other via:
\begin{eqnarray*}
&& \hat{\bmnabla} - \tilde{\bmnabla} = \bmS(\tilde{\bmb}), \\
&& \bmnabla - \tilde{\bmnabla} = \bmS(\Theta^{-1} \mathbf{d}\Theta),
\\
&& \hat{\bmnabla} - \bmnabla = \bmS(\bmf),
\end{eqnarray*}
where 
\[
\bmf \equiv \tilde{\bmb} - \Theta^{-1} \mathbf{d}\Theta,
\]
and $\bmS$ is the \emph{connection transition tensor} given in
component notation by
\begin{equation}
S_{ab}{}^{cd} \equiv \delta_a{}^c \delta_b{}^d + \delta_a{}^d
\delta_b{}^c - g_{ab}g^{cd}. 
\label{TransitionTensor}
\end{equation}

\medskip
In order to express tensorial objects we make use of a frame
$\{ \bme_\bma \}$ which is Weyl propagated along the congruence of
conformal geodesics with initial data given by
\eqref{ConformalGeodesicsData}. This frame is adapted to the
congruence by requiring $\bme_0 = \dot{\bmx}$. Moreover, it is also
required that initially
\[
\bmg_\star(\bme_\bma,\bme_\bmb)=\Theta^2_\star \tilde{\bmg} (\bme_\bma,\bme_\bmb) = \eta_{\bma\bmb}.
\] 
In view of the Weyl propagation condition $\hat{\bmnabla}_{\dot{\bmx}}
\bme_\bma =0$, it follows that the \emph{connection coefficients of
  the Weyl connection} $\hat{\Gamma}_\bma{}^\bmb{}_\bmc \equiv \langle
\bmsigma^\bmb, \nabla_\bma \bme_\bmc
\rangle $ satisfy the gauge condition
\[
\hat{\Gamma}_0{}^\bmb{}_\bmc =0.
\] 
Using the transformation rule between the connections and the properties of the Levi-Civita
connections it can be readily verified that $f_\bma=
\tfrac{1}{4}\hat{\Gamma}_\bma{}^\bmc{}_\bmc$ from where it follows
that
\[
f_0 =0.
\]
Finally, as a consequence of formulae
\eqref{WeylConnectionGaugeConditions} it follows that the components
of the Schouten tensor of the Weyl connection $\hat{\bmnabla}$ with
respect to the Weyl propagated frame satisfy
\[
\hat{L}_{0\bma}=0. 
\]

\subsubsection{Structure of the conformal boundary}
\label{Section:StructureConformalBoundary}

As it will be seen in Section
\ref{Section:ExistenceSolutionsCylinderSpatialInfinity} the radius
$a>0$ can be chosen small enough so that for each geodesic starting on
$p\in \mathcal{B}_a(i^0)$ with initial data of the form given by
\eqref{ConformalGeodesicsData}, there exists $\tau_\mathscr{I}(p)\geq 0$ such that
$\Theta(\pm\tau_\mathscr{I}(p))=0$ and the conformal geodesic does not contain
conjugate points in $[-\tau_\mathscr{I}(p),\tau_\mathscr{I}(p)]$. Moreover, it will be seen that if
$p=i^0$, then $\Theta=0$ for all $\tau$. Accordingly, for $p=i^0$ we
define $\tau_\mathscr{I}(i^0) \equiv \lim_{p\rightarrow i^0}
\tau_\mathscr{I}(p)$. In an abuse of notation, in what follows, the
lift of the point $p\in \mathcal{B}_a(i^0)$ to $\mathcal{C}_a$ will be
denoted, again, by $p$. Following these observations,
the domain on which we will be looking for solutions to the conformal
Einstein-Maxwell equations  ---see next section--- is of the form
\[
\mathcal{M}_{a,\kappa} \equiv \{  (\tau,p) \in \mathbb{R}\times
\mathcal{C}_{a,\kappa} \; | \;  -\tau_\mathscr{I}(p)\leq \tau \leq \tau_\mathscr{I}(p)
\}. 
\]

In a natural way, we define \emph{null infinity} as
$\mathscr{I}=\mathscr{I}^+\cup \mathscr{I}^-$ with
\[
\mathscr{I}^\pm \equiv \{  (\tau,p) \in \mathcal{M}_{a,\kappa} \; | \; \tau=\pm
\tau_\mathscr{I}, \; \rho(p) \neq 0\}.
\]
The \emph{cylinder at spatial infinity} is given by
\[
\cylinderinfinity \equiv \{ (\tau,p) \in \mathcal{M}_{a,\kappa} \; | \;
\rho(p) =0, \,  -\tau_\mathscr{I}(p) < \tau < \tau_\mathscr{I}(p) \}.
\]
Of interest are also the so-called \emph{critical sets}
\[
\cylinderinfinity^\pm \equiv \{ (\tau,p) \in \mathcal{M}_{a,\kappa} \; | \;
\rho(p) =0, \,  \tau =\pm \tau_\mathscr{I}(p) \},
\]
and
\[
\cylinderinfinity^0 \equiv \{ (\tau,p) \in \mathcal{M}_{a,\kappa} \; | \;
\rho(p) =0, \,  \tau =0 \},
\]
the intersection of $\mathcal{C}_{a,\kappa}$ with $\cylinderinfinity$. 

\medskip
Coordinates on $\mathcal{M}_{a,\kappa}$ are naturally dragged from
$\mathcal{C}_{a,\kappa}$ along the conformal geodesics. Similarly, the
vector fields $\bmpartial_\rho$, $\bmX_\pm$ extend in a unique way to
vectors on $\mathcal{M}_{a,\kappa}$ by requiring that they commute
with $\bmpartial_\tau$. 

\subsection{The extended conformal Einstein-Maxwell field equations}
\label{Section:CFE}

The conformal Einstein-Maxwell field equations have been
first considered in \cite{Fri91} where they have been used to show the
existence and stability of de Sitter-like spacetimes and the
semiglobal existence and stability of asymptotically Minkowskian
spacetimes. This formulation of the conformal field equations is
formulated in terms of geometric quantities associated to the
Levi-Civita connection of a conformally related (i.e. unphysical)
metric $\bmg$. In \cite{Fri95}, the vacuum conformal field equations have
been rewritten in terms of a (in principle arbitrary) Weyl connection
---we call these equations the \emph{extended conformal field
  equations}. The extended conformal field equations with matter have
been studied in \cite{LueVal12}, where applications to
the global existence of various electrovacuum spacetimes have been
considered.

\subsubsection{The frame formulation of the equations}
For completeness, we briefly review the general setting of the
extended conformal field equations. In the remainder of this
subsection, let $(\tilde{\mathcal{M}},\tilde{\bmg})$ denote a
spacetime satisfying the Einstein-Maxwell field equations
\eqref{EMFE1}-\eqref{EMFE3}. Let $\bmg$ denote the conformal metric
defined by the relation $\bmg = \Xi^2 \tilde{\bmg}$ where $\Xi$
denotes a (yet undetermined) conformal factor and let $\bmnabla$
denote its Levi-Civita connection. Let $\{\bme_\bma\}$,
$\bma=0,\ldots,3$ denote a frame field which is $\bmg$-orthogonal so
that $\bmg(\bme_\bma,\bme_\bmb) =\eta_{\bma\bmb}$, and let $\{
\bmomega^\bmb \}$ denote its dual cobasis ---i.e. $\langle
\bmomega^\bmb, \bme_\bma \rangle=\delta_\bma{}^\bmb$. As ${\bm \nabla}$
is the Levi-Civita connection of $\bmg$, one again has that its
connection coefficients, $\Gamma_\bma{}^\bmc{}_\bmb =\langle
\omega^\bmc, \nabla_\bma \bme_\bmb\rangle$ satisfy the usual metric
compatibility condition.

Let now $\hat{\bm \nabla}$ denote a Weyl connection constructed from
the Levi-Civita connection $\bm \nabla$ and a 1-form $\bmf$ using
formula $\hat{\bmnabla} -\bmnabla = \bmf$. If $\hat{\Gamma}_\bma{}^\bmc{}_\bmb =\langle
\omega^\bmc, \hat{\nabla}_\bma \bme_\bmb\rangle$ denotes the connection coefficients
of $\hat{\bm \nabla}$ with respect to the frame $\bme_\bma$, one has
then that
\begin{eqnarray*}
&& \hat{\Gamma}_\bma{}^\bmc{}_\bmb = \Gamma_\bma{}^\bmc{}_\bmb +
S_{\bma\bmb}{}^{\bmc\bmd}f_\bmd, \\
&& \phantom{\hat{\Gamma}_\bma{}^\bmc{}_\bmb}=  \Gamma_\bma{}^\bmc{}_\bmb + \delta _\bma{}^\bmc f_\bmb
+ \delta _\bmb{}^\bmc f_\bma -\eta_{\bma\bmb}\eta^{\bmc\bmd}f_\bmd.
\end{eqnarray*}
In particular, one has that $f_\bma =
\tfrac{1}{4}\hat{\Gamma}_\bma{}^\bmb{}_\bmb$, 
as $\Gamma_\bma{}^\bmb{}_\bmb=0$ in the case of a metric
connection.

Let $\hat{\Sigma}_\bma{}^\bmc{}_\bmb$ denote the torsion
of the connection $\hat{\bm\nabla}$. It is convenient to
distinguish between the expression for the components of the Riemann
tensor of the connection $\hat{\bm \nabla}$ in terms of the connection
coefficients $\hat{\Gamma}_\bma{}^\bmc{}_\bmb$ (the \emph{geometric
curvature} $\hat{P}^\bmc{}_{\bmd\bma\bmb}$) and the expression of the
Riemann tensor in terms of the Schouten and Weyl tensors
(the \emph{algebraic curvature}
$\hat{\rho}^\bmc{}_{\bmd\bma\bmb}$). Explicitly, one has that
\begin{eqnarray*}
&& \hat{P}^\bmc{}_{\bmd\bma\bmb} \equiv \bme_\bma(\hat{\Gamma}_\bmb{}^\bmc{}_\bmd) -
\bme_\bmb(\hat{\Gamma}_\bma{}^\bmc{}_\bmd) \\
&& \hspace{2cm} + \hat{\Gamma}_\bmf{}^\bmc{}_\bmd(
\hat{\Gamma}_\bmb{}^\bmf{}_\bma - \hat{\Gamma}_\bma{}^\bmf{}_\bmb ) +
\hat{\Gamma}_\bmb{}^\bmf{}_\bmd \hat{\Gamma}_\bma{}^\bmc{}_\bmf  -
\hat{\Gamma}_\bma{}^\bmf{}_\bmd \hat{\Gamma}_\bmb{}^\bmc{}_\bmf, \\
&& \hat{\rho}^\bmc{}_{\bmd\bma\bmb} \equiv \Xi d^\bmc{}_{\bmd\bma\bmb} + 2 S_{\bmd[\bma}{}^{\bmc\bme}\hat{L}_{\bmb]\bme}.
\end{eqnarray*}
Define the \emph{geometric zero quantities}
\begin{eqnarray*}
&& \hat{\Sigma}_\bma{}^\bmc{}_\bmb \equiv \langle \sigma^\bmc, [\bme_\bma,\bme_\bmb] \rangle
-(\hat{\Gamma}_\bma{}^\bmc{}_\bmb-\hat{\Gamma}_\bmb{}^\bmc{}_\bma) \\
&& \hat{\Xi}^\bmc{}_{\bmd\bma\bmb} \equiv \hat{P}^\bmc{}_{\bmd\bma\bmb} -
{\rho}^\bmc{}_{\bmd\bma\bmb}, \\
&& \hat{\Delta}_{\bmc\bmd\bmb} \equiv \hat{\nabla}_\bmc \hat{L}_{\bmd\bmb} - \hat{\nabla}_\bmd \hat{L}_{\bmc\bmb} - d_\bma d^\bma{}_{\bmb\bmc\bmd} - \Xi
T_{\bmc\bmd\bmb}, \\
&&  \Lambda_{\bmb\bmc\bmd}\equiv  \nabla_\bma
d^\bma{}_{\bmb\bmc\bmd} -
T_{\bmc\bmd\bmb},
\end{eqnarray*}
where $T_{\bmc\bmd\bmb} \equiv \nabla_{[\bmc} T_{\bmd]\bmb}$ 
and the \emph{Maxwell zero quantities}
\begin{eqnarray*}
&& M_\bma \equiv \nabla^\bmb F_{\bma\bmb}, \\
&& M_{\bma\bmb\bmc} \equiv \nabla_{[\bma}F_{\bmb\bmc]}.
\end{eqnarray*}
Then the \emph{extended conformal Einstein-Maxwell field equations} are given by the conditions
\begin{equation}
\hat{\Sigma}_\bma{}^\bmc{}_\bmb=0, \qquad \hat{\Xi}^\bmc{}_{\bmd\bma\bmb}=0,
\qquad \hat{\Delta}_{\bmc\bmd\bmb} =0, \qquad M_\bma =0, \qquad M_{\bma\bmb\bmc}=0.
\label{XCFEFrame}
\end{equation}
The fields $f_\bma$, $d_\bma$ and $\Xi$ are related to each other by
the constraint
\[
d_\bma = f_\bma + \nabla_\bma \Xi.
\]

The above conformal equations can be read as yielding differential conditions, respectively, for the frame
components $e_\bma{}^a$, the spin coefficients
$\hat{\Gamma}_\bma{}^\bmc{}_\bmb$ (including the 
the components $f_\bma$ of the 1-form $\bmf$), the components of the Schouten
tensor $\hat{L}_{\bma\bmb}$, the components of the rescaled Weyl
tensor $d^\bma{}_{\bmb\bmc\bmd}\equiv \Xi^{-1}
C^\bma{}_{\bmb\bmc\bmd}$, and the components of the unphysical Faraday tensor $F_{\bma\bmb}$. 

\medskip
\noindent
\textbf{Remark.} Equations \eqref{XCFEFrame} have to be supplemented with gauge
conditions or equations which determine the conformal factor $\Xi$ and
the 1-form $\bmd$. In the case of the particular applications to be
considered in this article, these fields will be determined by means
of a conformal Gaussian gauge fixed by the congruence of conformal
geodesics discussed in Section
\ref{Section:SpacetimeGaugeConsiderations}. Notice, however, that
other choices are possible ---e.g a gauge based on the properties of the
conformal curves discussed in \cite{LueVal12}.

\subsubsection{Spinorial formulation of the equations}
Hyperbolic reductions of the extended Einstein-Maxwell conformal field
equations \eqref{XCFEFrame} and a subsequent reduction to
spherical symmetry are best carried out using a \emph{space spinor
formalism} based on a spinor $\tau^{AA'}$ associated to a timelike
vector such that
$\tau_{AA'}\tau^{BA'} =\epsilon_A{}^B$. The spinorial counterparts of the fields
\[
\bme_\bma, \quad \hat{\Gamma}_\bma{}^\bmb{}_\bmc, \quad
\hat{L}_{\bma\bmb}, \quad d^\bma{}_{\bmb\bmc\bmd}, \quad F_{\bma\bmb}
\]
are given, respectively, by the spinor fields
\begin{equation}
\bme_{\bmA\bmB}, \quad \hat{\Gamma}_{\bmA\bmB\bmC\bmD}, \quad
\hat{L}_{\bmA\bmB\bmC\bmD}, \quad \phi_{\bmA\bmB\bmC\bmD}, \quad \varphi_{\bmA\bmB}.
\label{SpinorialUnknowns}
\end{equation}
Note that $\phi_{\bmA\bmB\bmC\bmD}=
\phi_{(\bmA\bmB\bmC\bmD)}$ and
$\varphi_{\bmA\bmB}=\varphi_{(\bmA\bmB)}$ so that they are already
decomposed into irreducible terms. For the fields
$\bme_{\bmA\bmB}$, $\hat{\Gamma}_{\bmA\bmB\bmC\bmD}$ and
$\hat{L}_{\bmA\bmB\bmC\bmD}$ one has the decompositions
\begin{eqnarray*}
&& \bme_{\bmA\bmB} = \tfrac{1}{2}\epsilon_{\bmA\bmB} \bme_Q{}^Q+
\bme_{(\bmA\bmB)}, \\
&& \hat{\Gamma}_{\bmA\bmB\bmC\bmD} = \tfrac{1}{\sqrt{2}} \big(
\xi_{\bmA\bmB\bmC\bmD}- \chi_{\bmA\bmB\bmC\bmD} \big) + \epsilon_{\bmA
  \bmC} f_{\bmD\bmB}, \\
&& \hat{L}_{\bmA\bmB\bmC\bmD} = \tfrac{1}{2} \epsilon_{\bmA\bmB}
\hat{L}_\bmQ{}^\bmQ{}_{\bmC\bmD}  + \tfrac{1}{2}\epsilon_{\bmC\bmD}
\hat{L}_{\bmA\bmB\bmQ}{}^\bmQ + \hat{L}_{(\bmA\bmB)(\bmC\bmD)},
\end{eqnarray*}
where $f_{\bmA\bmB}$ is the space spinor counterpart of the field
$f_\bma$. It satisfies, in turn, the split
\[
f_{\bmA\bmB} = \tfrac{1}{2}\epsilon_{\bmA\bmB} f_\bmQ{}^\bmQ + f_{(\bmA\bmB)}.
\]
The spinor fields $\xi_{\bmA\bmB\bmC\bmD}$ and
$\chi_{\bmA\bmB\bmC\bmD}$ correspond, at least on an initial
hypersurface, to the spinorial counterparts of the intrinsic connection
to the hypersurface and the extrinsic curvature, respectively. It is
recalled that in the space spinor formalism the trace part of
spinorial fields corresponds to the time components of tensors ---see
e.g. \cite{Som80}.  

\medskip
Expressions of the spinorial version of the Einstein-Maxwell equations
\eqref{XCFEFrame} in terms of the fields \eqref{SpinorialUnknowns} can
be given in terms of the following spinor zero-quantities:
\begin{equation}
\hat{\Sigma}_{\bmA\bmB\bmC\bmD} =0, \qquad
\hat{\Xi}_{\bmA\bmB\bmC\bmD\bmE\bmF}=0 \qquad
\hat{\Delta}_{\bmA\bmB\bmC\bmD}=0, \qquad 
\hat{\Lambda}_{\bmA\bmB\bmC\bmD}=0, \qquad \hat{M}_{\bmA\bmB} =0.
\label{SpinorialXCFE}
\end{equation}
The explicit form of these zero-quantities will not be given here. The
interested reader is referred to \cite{LueVal12}. 

\subsubsection{The hyperbolic reduction of the spinorial conformal Einstein-Maxwell
  field equations}
In the remainder of this article we will consider evolution equations
for the various spinorial conformal fields obtained through an
hyperbolic reduction procedure based on the gauge properties of
conformal geodesics discussed in Section
\ref{Section:WeylGaugeConditions}. In view of our particular
application, the various spinor fields and their corresponding tensors will be
lifted to the bundle space $\mathcal{M}_{a,\kappa}$. In a slight
abuse of notation, we will denote the spinorial fields and their lifts
to the bundle space with the same symbol. The precise nature of the object
should be clear from the context. 

\medskip
The gauge conditions discussed in Section
\ref{Section:WeylGaugeConditions}, expressed in terms of spinorial
objects, take the form
\begin{equation}
\bme_\bmQ{}^\bmQ = \sqrt{2} \bmpartial_\tau, \qquad
\hat{\Gamma}_\bmQ{}^\bmQ{}_{\bmC\bmD}=0, \qquad f_\bmQ{}^\bmQ =0,
\qquad \hat{L}_\bmQ{}^\bmQ{}_{\bmC\bmD}=0,
\label{SpinorialGaugeConditions}
\end{equation}
on $\mathcal{M}_{a,\kappa}$. Hence, in particular, one has that 
\[
\bme_{\bmA\bmB} =\sqrt{2} \epsilon_{\bmA\bmB} \bmpartial_\tau +
\big( e^0_{\bmA\bmB} \bmpartial_\tau + e^1_{\bmA\bmB} \bmpartial_\rho
+ e^+_{\bmA\bmB} \bmX_+ + e^-_{\bmA\bmB} \bmX_- \big),
\]
where $e^\mu_{\bmA\bmB} = e^\mu_{(\bmA\bmB)}$ with $\mu=0, \, 1, \, +,
\, -$. 

\medskip
Suitable evolution equations are obtained from the following
components of \eqref{SpinorialXCFE}
\begin{equation}
\hat{\Sigma}_\bmQ{}^\bmQ{}_{\bmC\bmD} =0, \qquad
\hat{\Xi}_\bmQ{}^\bmQ{}_{\bmC\bmD\bmE\bmF}=0, \qquad
\hat{\Delta}_\bmQ{}^\bmQ{}_{\bmC\bmD} =0, \qquad
\hat{\Lambda}_{(\bmA\bmB\bmC\bmD)} =0, \qquad \hat{M}_{(\bmA\bmB)}=0,
\label{SpinorialEvolutionEquations}
\end{equation}
together with the gauge conditions
\eqref{SpinorialGaugeConditions}. The above evolution equations
satisfy a suitable \emph{propagation of the constraints}
result. Namely, if equations \eqref{SpinorialXCFE} are satisfied
initial on some hypersurface then they are also satisfied in the
domain of dependence as long as the evolution equation
\eqref{SpinorialEvolutionEquations} hold ---see \cite{LueVal12} for
more details. 

\subsubsection{The evolution equations in spherical symmetry}
\label{Section:SSEvolutionEquations}

A spherical symmetry reduction of the evolution equations
\eqref{SpinorialEvolutionEquations} can be implemented by
expressing the various quantities in terms of the spinors $\epsilon_{\bmA\bmB}$, $x_{\bmA\bmB}$,
$y_{\bmA\bmB}$ and $z_{\bmA\bmB}$ introduced in Section
\ref{Section:LiftsBundleSpace} and then making an Ansatz based on the
spin-weight of the components. Save for the case of frame coefficients
$e_{\bmA\bmB}^\pm$, only components with spin-weight zero are
considered. A computation shows that under these circumstances a
suitable Ansatz for an Einstein-Maxwell field in the present formalism
is given by 
\begin{eqnarray*}
&& e^0_{\bmA\bmB} = e^0 x_{\bmA\bmB}, \qquad e^1_{\bmA\bmB}
=e^1x_{\bmA\bmB}, \qquad e^+_{\bmA\bmB} = e^+ z_{\bmA\bmB}, \qquad
e^-_{\bmA\bmB} = e^- y_{\bmA\bmB}, \\
&& f_{\bmA\bmB} = f x_{\bmA\bmB}, \qquad \xi_{\bmA\bmB\bmC\bmD} =
\tfrac{1}{\sqrt{2}}\xi (\epsilon_{\bmA\bmC} x_{\bmB\bmD} +
\epsilon_{\bmB\bmD} x_{\bmA\bmC}), \\
&& \chi_{\bmA\bmB\bmC\bmD} = \chi_2 \epsilon^2_{\bmA\bmB\bmC\bmD} +
\tfrac{1}{3}\chi_h h_{\bmA\bmB\bmC\bmD}, \\
&& \hat{L}_{\bmA\bmB\bmC\bmD} = \theta_2 \epsilon^2_{\bmA\bmB\bmC\bmD} +
\tfrac{1}{3}\theta_h h_{\bmA\bmB\bmC\bmD}  + \tfrac{1}{\sqrt{2}}\theta_x\epsilon_{\bmC\bmD}
x_{\bmA\bmB}, \\
&& \phi_{\bmA\bmB\bmC\bmD} = \phi \epsilon^2_{\bmA\bmB\bmC\bmD},
\qquad \varphi_{\bmA\bmB} =\varphi x_{\bmA\bmB}.
\end{eqnarray*}
where all the coefficients in the above Ansatz are real. For more details on the motivations behind the above
spherical symmetric Ansatz, we refer the reader to \cite{Fri98a,Val12}. A lengthy computation
using the suite {\tt xAct} for tensorial
and spinorial manipulations for {\tt Mathematica} ---see \cite{xAct,GarMar12}--- yields the
following \emph{spherically symmetric conformal
evolution equations}
\begin{subequations}
\begin{eqnarray}
&& \dot{e}^0 = \tfrac{1}{3}(\chi_2-\chi_h) e^0 -f, \label{Reduced:SSEe0}\\
&& \dot{e}^1 = \tfrac{1}{3}(\chi_2-\chi_h)e^1, \label{Reduced:SSEe1}\\
&& \dot{e}^\pm = -\tfrac{1}{6}(\chi_2 +2
\chi_h)e^\pm, \label{Reduced:SSEepm} \\
&& \dot{f} = \tfrac{1}{3} (\chi_2 - \chi_h) f + \theta_x, \label{Reduced:SSEf}\\
&& \dot{\xi}_x = -\tfrac{1}{6}(\chi_2 + 2\chi_h) \xi_x-\tfrac{1}{2}\chi_2 f - \theta_x, \label{Reduced:SSExix}\\
&& \dot{\chi}_2 = \tfrac{1}{6} (\chi_2 -4\chi_h)\chi_h -\theta_2 +\Theta\phi, \label{Reduced:SSEchi2}\\
&& \dot{\chi}_h = -\tfrac{1}{6} \chi^2_2 - \tfrac{1}{3}\chi^2_h -\theta_h, \label{Reduced:SSEchih}\\
&& \dot{\theta}_x=\tfrac{1}{3}(\chi_2-\chi_h)\theta_x
-\tfrac{1}{3}d_x \phi +\tfrac{1}{2}\Theta^2 d_x \varphi^2 -\tfrac{1}{4} \Theta^2\varphi^2 f, \label{Reduced:SSEthetax}\\
&& \dot{\theta}_2 = \tfrac{1}{6}(\chi_2-2\chi_h)\theta_2
-\tfrac{1}{3}\chi_2 \theta_h -
\phi \dot{\Theta}+ \tfrac{1}{4}\Theta^2 \varphi^2(3\chi_2+4\chi_h)-\Theta \dot{\Theta} \varphi^2, \label{Reduced:SSEtheta2}\\
&& \dot{\theta}_h =-\tfrac{1}{6}\chi_2 \theta_2 - \tfrac{1}{3}
\chi_h \theta_h- \Theta \dot{\Theta}\varphi^2 +\tfrac{1}{4} \Theta^2
\chi_h \varphi^2, \label{Reduced:SSEthetah}\\
&& \dot{\phi}=
-\tfrac{1}{2}(\chi_2+2\chi_h)\phi+\tfrac{1}{2}(\chi_2 +2
\chi_h )\Theta \varphi^2 -\dot{\Theta}\varphi^2, \label{Reduced:SSEphi} \\
&& \dot{\varphi} = -\tfrac{1}{3}\left( \chi_2 + 2\chi_h\right)
\varphi, \label{Reduced:SSEvarphi}
\end{eqnarray}
\end{subequations}
where, as before, $\dot{\phantom{x}}$ denotes differentiation with respect to the time
coordinate $\tau$. Notice that in the conformal Gaussian gauge used to
write the above equations, the field unknowns associated to a
spherically symmetric electrovacuum field are dynamic ---that is, the
evolution in this gauge does not follow the static Killing vector of
the spacetime. Note however, that the evolution equations are mere
\emph{transport equations} (i.e. ordinary differential equations)
along the congruence of conformal geodesics ---the evolution along a
given conformal geodesic is decoupled from the evolution in nearby
curves. This decoupling disappears when considering other (spherically
symmetric) systems like the Einstein-Yang-Mills or the
Einstein-conformally invariant wave equation ---see e.g. \cite{LueVal13b}. 

\medskip
The essential dynamics of the evolution equations
\eqref{Reduced:SSEe0}-\eqref{Reduced:SSEvarphi} is steered by a subset
thereof which decouples from the
rest of the system ---we call this subsystem the \emph{core system}. Letting
\[
X \equiv \chi_h + \tfrac{1}{2}\chi_2, \qquad L\equiv \theta_h +
\tfrac{1}{2} \theta_2,
\]
the core system is given by the equations
\begin{subequations}
\begin{eqnarray}
&& \partial_\tau L=-\tfrac{1}{3} XL -\tfrac{1}{2} \dot{\Theta}
\phi+\tfrac{3}{4}\Theta^2 X\varphi^2 - \tfrac{3}{2}\Theta
\dot{\Theta} \varphi^2 , \label{Reduced:P} \\
&& \partial_\tau X=-\tfrac{1}{3} X^2 -L + \tfrac{1}{2} \Theta \phi, \label{Reduced:X}\\
&& \partial_\tau \phi = -X\phi +\Theta X \varphi^2 -\dot{\Theta}\varphi^2 \label{Reduced:phi}, \\
&& \partial_\tau \varphi = -\tfrac{2}{3} X \varphi.\label{Reduced:varphi}
\end{eqnarray}
\end{subequations}

\medskip
The evolution equations
\eqref{Reduced:SSEe0}-\eqref{Reduced:SSEvarphi} ---or
\eqref{Reduced:P}-\eqref{Reduced:varphi}--- are supplemented by the evolution equations for the
fields $\Theta$, $d$, $d_x$ provided by the conformal Gaussian
gauge as discussed in Section
\ref{Section:CanonicalConformalFactor}. Recalling that the
spinorial counterpart of the unphysical energy-momentum tensor $\bmT$
is given by the spinor $T_{AA'BB'} = \varphi_{AB}
\bar{\varphi}_{A'B'}$, it follows that
\[
\bmT(\dot{\bmx},\dot{\bmx}) = \tfrac{1}{2} \varphi_{AB}
\bar{\varphi}_{A'B'} \tau^{AA'}\tau^{BB'} = \tfrac{1}{2}\varphi^2.
\]
Note the overal factor in the expression arises from the
normalisation $\tau_{AA'} \tau^{AA'}=2$. The spinorial counterpart of $T_{\bmzero\bmi}$ is given by
\begin{equation}
T_{AA'B'(B}\tau^{AA'} \tau_{C)}{}^{B'} = \varphi_{A(B}
\varphi^{\dagger A}{}_{C)} = \varphi^2 x_{A(B} x^A{}_{C)} =0. 
\label{FluxVectorSphericalSymmetry}
\end{equation}
It then follows that the third order evolution equation for the
canonical conformal factor $\Theta$
---cf. \eqref{ThetaEvolutionEquation3ALT}--- can be written as 
\begin{equation}
\dddot{\Theta} = \tfrac{1}{2}\Theta^2 \dot{\Theta} \varphi^2
-\tfrac{1}{3} X \Theta^3 \varphi^2. 
\label{Reduced:SSETheta}
\end{equation}
If 
\[
d_{\bmA\bmB} = \tfrac{1}{2} \epsilon_{\bmA\bmB} d + d_{(\bmA\bmB)}
\]
 denotes the space-spinor counterpart of the field $d_\bma$, it
 follows from \eqref{Equation:d0}
that
\begin{equation}
d = \dot{\Theta},
\label{Reduced:SSEd0}
\end{equation}
while from \eqref{Equation:di} together with
\eqref{FluxVectorSphericalSymmetry} one finds that $d_{(\bmA\bmB)} =
d_x x_{\bmA\bmB}$ is constant along the conformal geodesics and can be
specified directly from initial data ---see equation \eqref{dxData} in
Section \ref{Section:InitialData} below.  

\medskip
\noindent
\textbf{Remark 1.} Equations
\eqref{Reduced:SSEe0}-\eqref{Reduced:SSEvarphi} together with
\eqref{Reduced:SSETheta} and \eqref{Reduced:SSEd0} are the complete
evolution system for spherically symmetric electrovacuum
spacetimes. Strictly speaking, in order to obtain a system which is
purely of first order, one should introduce a further variable
representing the second order derivative of the conformal factor. The hyperbolic reduction can be achieving
without introducing the derivative of the Maxwell spinor
$\varphi_{AB}$ as a further unknown ---see \cite{Fri91,LueVal12}. This
is feature is a consequence of the spherical symmetry where
$\nabla_{AA'}\varphi_{BC}$ has only two components ---the time and the
radial ones. The time derivative is readily available from the
evolution equations while the radial one can be obtained from the
constraint implied by the Maxwell equations ---see \cite{Val12} for
more details.

\medskip
\noindent
\textbf{Remark 2.} Given a solution to equations
\eqref{Reduced:SSEe0}-\eqref{Reduced:SSEvarphi} together with
\eqref{Reduced:SSETheta} and \eqref{Reduced:SSEd0}, one can construct
a metric tensor $\bmg$ on $\mathcal{M}_{\bma,\kappa}$ using the
decomposition
\begin{equation}
\bmg =\mathbf{d}\tau \otimes \mathbf{d}\tau -
\frac{e^0_x}{e^1_x}\left( \mathbf{d}\tau \otimes \mathbf{d}\rho +
  \mathbf{d}\rho \otimes \mathbf{d}\tau \right) - \left(
  \frac{1}{(e^1_x)^2} -\left( \frac{e^0_x}{e^1_x}\right)^2 \right)
\mathbf{d}\rho \otimes \mathbf{d}\rho -\frac{1}{e^+_z
  e^-_y}\bmsigma.
\label{SpacetimeMetric}
\end{equation}
as long as $e^1_x, \, e^+_z
  e^-_y\neq 0$ ---see \cite{Fri98a,Val12} for further details. This
  metric on a bundle space can be projected down to a (unphysical)
  spacetime metric which will be denoted, again, by $\bmg$.

\subsubsection{Initial data for the evolution equations}
\label{Section:InitialData}

Initial data for the evolution equations
\eqref{Reduced:SSEe0}-\eqref{Reduced:SSEvarphi},
\eqref{Reduced:SSETheta} and \eqref{Reduced:SSEd0} on
the extended bundle space $\mathcal{C}_{a,\kappa}$ is obtained
following the discussion of Section \ref{Section:ExtendedBundleSpace}
and observing the expressions in Section \ref{Section:SolderForms}. One finds that
\begin{subequations}
\begin{eqnarray}
& e^0_{\bmA\bmB}=0, \quad e^1{}_{\bmA\bmB}=\rho x_{\bmA\bmB}, \quad e^-_{\bmA\bmB} =y_{\bmA\bmB}
\quad e^+_{\bmA\bmB} = z_{\bmA\bmB},& \label{GeometricData1}\\
& f = f_{\bmA\bmB} x^{\bmA\bmB} =1 , \quad \xi_{\bmA\bmB\bmC\bmD} =0, \quad \chi_{\bmA\bmB\bmC\bmD} =0,& \\
& \Theta_{\bmA\bmB\bmC\bmD} =\frac{\displaystyle 6m\rho}{\displaystyle
  (1+m\rho)^2}\epsilon^2_{\bmA\bmB\bmC\bmD} , \quad \phi_{\bmA\bmB\bmC\bmD} =-
\frac{\displaystyle 6m}{\displaystyle 1+m\rho}\epsilon^2_{\bmA\bmB\bmC\bmD}, \quad
\varphi_{\bmA\bmB} =-m x_{\bmA\bmB}.& \label{GeometricData2}
\end{eqnarray}
\end{subequations}
The initial data for the conformal gauge unknowns is given by
\begin{subequations}
\begin{eqnarray}
& \Theta_\star = \frac{\displaystyle \rho}{\displaystyle 1+m\rho}, \qquad \dot{\Theta}_\star =0, \qquad  \ddot{\Theta}_\star = -\frac{\displaystyle \rho(2+m\rho)^2}{\displaystyle 2(1+m\rho)^3}, & \label{GaugeData1}\\
& d_{x\star} = \frac{\displaystyle 2\rho+m\rho^2}{\displaystyle (1+m\rho)^2}. & \label{dxData}
\end{eqnarray}
\end{subequations}
One observes that, in particular, $\Theta_\star = \dot{\Theta}_\star =
\ddot{\Theta}_\star =0$ at $\rho=0$. 

\subsubsection{Conformal deviation equations}
The gauge used in the construction of the cylinder at
spatial infinity hinges on the properties of conformal
geodesics. Hence it is important to verify that the congruence of curves is
free of conjugate points on $\mathcal{M}_{a,\kappa}$.

\medskip
In what follows, let $\eta_a$ denote the conformal Jacobi field
measuring the deviation of the curves in the congruence of conformal
geodesics. Its spinorial counterpart has the space-spinor
representation
\[
\eta_{\bmA\bmB} = \tfrac{1}{2}\eta \epsilon_{\bmA\bmB} + \eta_{(\bmA\bmB)},
\]
where $\eta\equiv \eta_\bmQ{}^\bmQ$. In \cite{LueVal09} it has been
shown that $\eta$, $\eta_{(\bmA\bmB)}$ satisfy the evolution
equations
\begin{eqnarray*}
&& \partial_\tau \eta = f_{(\bmA\bmB)} \eta^{\bmA\bmB}, \\
&& \partial_\tau \eta_{(\bmA\bmB)} =
\chi_{(\bmC\bmD)(\bmA\bmB)} \eta^{(\bmC\bmD)}. 
\end{eqnarray*}
Conjugate points in the congruence arise if $\eta_{(\bmA\bmB)}=0$. In
view of the spherical symmetry of the setting one has that 
\[
\eta_{(\bmA\bmB)} = \eta_x x_{\bmA\bmB}. 
\]
Hence, the evolution equations for $\eta$ and $\eta_{(\bmA\bmB)}$
reduce to
\begin{subequations}
\begin{eqnarray}
&&\partial_\tau \eta = -f \eta_x, \label{SSDeviation1}\\
&&\partial_\tau \eta_x = \tfrac{1}{3}( \chi_h -\chi_2) \eta_x. \label{SSDeviation2}
\end{eqnarray}
\end{subequations}
Without loss of generality, initial data for the above fields can be
set to be 
\[
\eta_\star =0, \qquad \eta_{x\star} =1 \qquad \mbox{on} \quad \mathcal{C}_a.
\]

\subsection{Existence of solutions on $\mathcal{M}_{a,\kappa}$}
\label{Section:ExistenceSolutionsCylinderSpatialInfinity}

The evolution equations
\eqref{Reduced:SSEe0}-\eqref{Reduced:SSEvarphi},
\eqref{Reduced:SSETheta} and \eqref{Reduced:SSEd0} on the bundle
manifold $\mathcal{M}_{a,\kappa}$ together with the initial data
\eqref{GeometricData1}-\eqref{GeometricData2} and
\eqref{GaugeData1}-\eqref{dxData} on $\mathcal{C}_{a,\kappa}$ are all regular for
sufficiently small $a>0$. Hence they give rise to a \emph{regular finite initial
  value problem near spatial infinity}. In view of the particular form of the
equations, standard theory ensures the existence of a local solution
in a neighbourhood of $\mathcal{C}_{a,\kappa}$. Naturally, we are
interested in extending the existence result to the whole of
$\mathcal{M}_{a,\kappa}$. This requires a slightly more detailed
analysis. 

\medskip
Letting 
\[
\mathbf{u} \equiv \big( e^0, \, e^1, e^\pm, f, \xi_x, \chi_2 , \chi_h,
\theta_x, \theta_2, \theta_h, \phi, \varphi, \Theta, \Sigma, \Lambda\big)
\]
with $\Sigma \equiv \dot{\Theta}$, $\Lambda\equiv \ddot{\Theta}$, one
can rewrite the initial value problem in the form
\begin{equation}
\partial_\tau \mathbf{u} = F(\mathbf{u},\tau,\rho;m), \qquad
\mathbf{u}(0,\rho;m)=\mathbf{u}_\star(\rho;m),
\label{EvolutionSystem}
\end{equation}
where $F$ and $\mathbf{u}_\star$ are analytic functions of their
arguments. The particular case $m=0$ (i.e. the Minkowski spacetime)
can be solved explicitly with the only non-vanishing geometric fields given by 
\[
e^0 =-\tau, \qquad e^1 =\rho, \qquad e^\pm =1, \qquad f=1,
\]
while the fields associated to the conformal gauge are
\[
\Theta = \rho( 1-\tau^2), \qquad d_{\bmA\bmB} =2\rho x_{\bmA\bmB}.  
\]
Consequently, this solution exists for all
$\tau,\,\rho\in \mathbb{R}$. Moreover, from
\eqref{SSDeviation1}-\eqref{SSDeviation2} it follows that
\[
\eta = -\tau, \qquad \eta_x =1
\]
so that no conjugate points arise in the congruence of conformal
geodesics if $m=0$. 

\medskip
 Now, returning to the case $m\neq 0$, by Cauchy stability of ordinary
differential equations ---see e.g. \cite{Har87}--- given  $\tau_\bullet>1$ there exist
$m_\bullet>0$, $\rho_\bullet>0$ such that the solution
$\mathbf{u}(\tau,\rho;m)$ is analytic in all variables and exists for
\[
|\tau| \leq \tau_\bullet, \qquad \rho\leq \rho_\bullet, \qquad |m|\leq m_\bullet.
\]
By choosing $\tau_\bullet$ sufficiently large and observing the
properties of the reference $m=0$ solution,  one can ensure that for
each conformal geodesic  with $0<\rho<\rho_\bullet$ there exists a
$\tau_\mathscr{I}<\tau_\bullet$ such that $\Theta|_{\pm \tau_\mathscr{I}}=0$,
$\mathbf{d}\Theta|_{\pm \tau_\mathscr{I}}\neq 0$. In order to obtain a result that is valid for any value of $m$, it is
noticed that equations
\eqref{Reduced:SSEe0}-\eqref{Reduced:SSEvarphi},
\eqref{Reduced:SSETheta} and \eqref{Reduced:SSEd0} together with the
data \eqref{GeometricData1}-\eqref{GeometricData2},
\eqref{GaugeData1}-\eqref{dxData} together with equations are invariant under the rescaling
\[
 m \mapsto \frac{\displaystyle 1}{\displaystyle \lambda}m, \qquad \rho\mapsto \lambda \rho, \qquad \phi \mapsto \frac{\displaystyle 1}{\displaystyle \lambda} \phi, \qquad \varphi \mapsto \frac{\displaystyle 1}{\displaystyle \lambda} \varphi, \qquad e^1 \mapsto \lambda e^1, \qquad \Theta \mapsto\lambda \Theta,
\]
for $\lambda>0$. Consequently, for any arbitrary $m$ it is always
possible to obtain a solution to the system \eqref{EvolutionSystem}
reaching null infinity if $a$, and hence $\rho$, is sufficiently small. By a similar
argument based on Cauchy stability, one can conclude that if $\rho$ is
sufficiently small, the congruence of conformal geodesics in $\mathcal{M}_{a,\kappa}$ is free of conjugate points. Once a
solution to the conformal evolution equations has been obtained, the existence of a
solution to the Einstein constraint equations follows from the
analysis of the propagation of the constraints ---see
\cite{Fri95,LueVal12}. In conjunction with the result on the
propagation of the constraints given in \cite{LueVal12} one obtains the following:

\begin{theorem}
Given $m>0$, there exists $a>0$ such that on $\mathcal{M}_{a,\kappa}$
there exists a unique smooth solution to the spherically symmetric
initial value problem at spatial infinity for the extended conformal
Einstein-Maxwell equations. For each $\rho\in [0,a]$ there exists
$\tau_\mathscr{I}(\rho)>0$ such that $\Theta|_{\pm \tau_\mathscr{I}(\rho)}=0$. For $\rho
\neq 0$, the sets
\[
\mathscr{I}^\pm \equiv \{ p \in \mathcal{M}_{a,\kappa} \; | \; \Theta(p) =0   \}
\]
correspond to the future and past null infinities of the metric
$\Theta^{-2} \bmg$ where the metric $\bmg$ is given by expression
\eqref{SpacetimeMetric}.  
\end{theorem}

\medskip
\noindent
\textbf{Remark.} It is important to emphasise, that the above result
is independent of the sign of $m$. 

\begin{figure}[t]
\begin{center}
\includegraphics[width=0.25\textwidth]{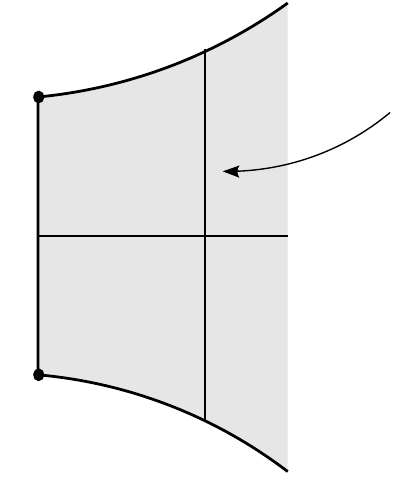}
\put(-80,110){$\mathscr{I}^+$}
\put(-80,10){$\mathscr{I}^-$}
\put(-80,50){$\mathcal{S}$}
\put(-110,96){$\cylinderinfinity^+$}
\put(-110,16){$\cylinderinfinity^-$}
\put(-110,60){$\cylinderinfinity$}
\put(0,95){\emph{conformal geodesic}}
\end{center}
\caption{Schematic 2-dimensional diagram of the bundle manifold
  $\mathcal{M}_{a,\kappa}$ giving rise to the cylinder at
  spatial infinity. The null hypersurfaces $\mathscr{I}^\pm$ correspond to the
  horizon, $\cylinderinfinity$ is the cylinder at  spatial infinity proper
  and $\cylinderinfinity^\pm$ are the critical sets where the cylinder
meets null infinity. The conformal geodesics on which this construction
is based correspond to straight vertical lines as indicated. Notice
that the diagram is not conformal ---null hypersurfaces are not
straight lines with slope of  $45^\circ$ or $135^\circ$. }
\label{Figure:CylinderSpatialInfinity}
\end{figure}
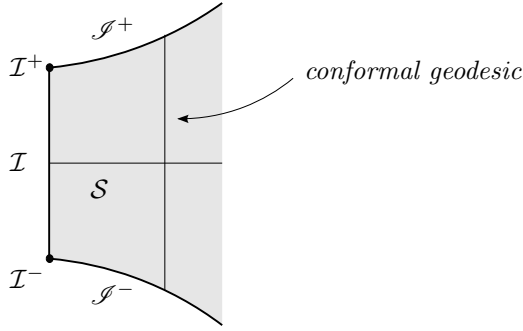

\subsubsection{The solution on the cylinder at spatial infinity}

On the conformal geodesics with $\rho=0$, the evolution equations
\eqref{Reduced:SSEe0}-\eqref{Reduced:SSEvarphi},
\eqref{Reduced:SSETheta} and \eqref{Reduced:SSEd0} simplify and it is
possible to obtain the solution in closed form.  If $\rho=0$, the data
for the conformal factor satisfies
\[
\Theta_\star = \dot{\Theta}_\star = \ddot{\Theta}_\star =0.
\]
It follows then from the evolution equation for $\Theta$ that
\[
\Theta=\dot{\Theta}=0
\]
for all later times.  As a consequence of this observation, and taking into account
the initial data for the various fields one finds that the only
non-vanishing  unknowns are given by
\begin{equation}
e^0=-\tau, \qquad e^\pm =1, \qquad f=1, \qquad \phi=-6m, \qquad \varphi=-m.
\label{SolutionCylinderSpatialInfinity}
\end{equation}

\section{The cylinders at the horizon}
\label{Section:CylindersHorizon}

The purpose of this section is to show that it is possible to obtain
regular representations of the points $i^+$ and $c^0$ ---see Figure
\ref{ConformalDiagrameRN}.  This achieved by combining the
construction of the cylinder at spatial infinity discussed in the
previous section with the conformal isometry of the extremal
Reissner-Nordstr\"om spacetime. We call the resulting constructions
the \emph{cylinders at the horizon}. As it will be seen the key
features of the construction are independent of the sign of $m$. In
order to distinguish quantities near the horizon from quantities near
conformalboundary we will use an overbar for all variables related to
the constructions of the \emph{cylinders at the horizon}.

\subsection{The bundle manifolds $\bar{\mathcal{C}}_a(i^+)$ and $\bar{\mathcal{C}}_a(c^0)$} 
In what follows, let $\bar{\i}$ indicate either $i^+$ or $c^0$ and let
$(\mathcal{S}_{\bar{\i}}, -\bar{\bmdelta})$ denote the conformal extensions of
$(\tilde{\regionIIIslices},\tilde{\bmh})$ or $(\tilde{\mathcal{S}}, \tilde{\bmh})$ 
 as discussed in Section 
\ref{Section:BasicConformalExtensions} and
\[
\bar{\bmdelta} = \mathbf{d}r \otimes \mathbf{d}r + r^2\bmsigma.
\]
By analogy to
the construction of the cylinder at spatial infinity, we begin by
considering the blow up of the point $\bar{\i}$. Given $\bar{a}>0$ and the open
ball $\mathcal{B}_{\bar{a}}(\bar{\i})\subset \mathcal{S}_{\bar{\i}}$ one defines, in analogy to the definition of
the bundle space $\mathcal{C}_a$ given in equation
\eqref{DefinitionCa}, the bundle manifold
\[
\bar{\mathcal{C}}_{\bar{a}}\equiv \big\{ \delta_\bmA(t,r) \in
SU(\mathcal{B}_{\bar{a}}(\bar{\i})) \; | \; 0\leq r <{\bar{a}}\big\},
\]
where $SU(\mathcal{B}_{\bar{a}}(\bar{\i}))$ is the bundle of normalised spin frames
over $\mathcal{B}_{\bar{a}}(\bar{\i})$ with structure group $SU(2,\mathbb{C})$, and
$\delta_\bmA(t,r)$ correspond to the frames such that the vector $\bme_3(t,r)=
\sigma_3{}^{\bmA\bmB} \delta_\bmA(t,r) \delta_\bmB(t,r)$ is tangent to the
radial geodesics with affine parameter $r$.

\medskip
Spinor fields on $\mathcal{B}_{\bar{a}}(\bar{\i})$ can be lifted to the bundle space
$\bar{\mathcal{C}}_{\bar{a}}$ in an analogous way to the lifts to
$\mathcal{C}_a$. Again, the resulting fields can be spanned in terms
of the basic spinors $x_{\bmA\bmB}$, $y_{\bmA\bmB}$, $z_{\bmA\bmB}$
and $\epsilon_{\bmA\bmB}$. Moreover, the vector fields $\{
\bmpartial_r, \, \bmX,\, \bmX_+, \, \bmX_- \}$ with $\bmX,\, \bmX_+,
\, \bmX_-$ the basis of the Lie algebra $\mathfrak{su}(2,\mathbb{C})$
discussed in Section \ref{Section:VectorFieldsCa} constitute a set
of frame fields on $\bar{\mathcal{C}}_{\bar{a}}$. As in Section
\ref{Section:VectorFieldsCa}, let $\bmalpha,\, \bmalpha^+,\, \bmalpha^-$
denote the dual 1-forms to  $\bmX,\, \bmX_+,
\, \bmX_-$. 

\medskip
Frame fields and solder forms on $\bar{\mathcal{C}}_{\bar{a}}$ are introduced
in a natural manner via
\begin{eqnarray*}
&& \bar{\bme}_{\bmA\bmB} = x_{\bmA\bmB} \bmpartial_r + \frac{1}{r}
z_{\bmA\bmB} \bmX_+ + \frac{1}{r} y_{\bmA\bmB} \bmX_, \\
&& \bar{\bmsigma}^{\bmA\bmB} = -x^{\bmA\bmB} \mathbf{d} r -2r y^{\bmA\bmB}
\bmalpha^+ - 2r z^{\bmA\bmB} \bmalpha^-,
\end{eqnarray*}
so that 
\[
\bar{\bmh} = h_{\bmA\bmB\bmC\bmD} \bar{\bmsigma}^{\bmA\bmB} \otimes
\bar{\bmsigma}^{\bmC\bmD}, \qquad \langle \bar{\bmsigma}^{\bmA\bmB},
\bar{\bme}_{\bmC\bmD}\rangle = h^{\bmA\bmB}{}_{\bmC\bmD},
\]
where $\bar{\bmh}$ denotes (again, in an abuse of notation) the lift to  $\bar{\mathcal{C}}_{\bar{a}}$ of the conformal
metric $\bar{\bmh}=-\bar{\bmdelta}$. The spin coefficients are given by
\[
\bar{\gamma}_{\bmA\bmB\bmC\bmD} = \frac{1}{2r}(\epsilon_{\bmA\bmC}
x_{\bmB\bmD} + \epsilon_{\bmB\bmD} x_{\bmA\bmC}). 
\]

\medskip
In analogy to the discussion of Section
\ref{Section:ExtendedBundleSpace}, given a non-negative smooth
function $\bar{\kappa}$ we introduce the \emph{extended bundle space}
\[
\bar{\mathcal{C}}_{\bar{a},\bar{\kappa}} \equiv \{ \bar{\kappa}^{1/2} \delta_\bmA \; | \; \delta_\bmA
\in \bar{\mathcal{C}}_{\bar{a}} \}. 
\]
In what follows $\bar{\phi}_{ABCD}$, $\bar{\Theta}_{ABCD}$ and $\bar{\varphi}_{AB}$ will now
denote the spinorial counterparts of the fields $d_{\bmi\bmj}$,
$\hat{L}_{\bmi\bmj}$ and $E_\bmi$ on $\mathcal{S}_{\bar{\i}}$ as
discussed in Section \ref{Section:ConformalConstraints}. Following our
standard usage,  we denote their lifts to the bundle space
$\bar{\mathcal{C}}_a$ by $\bar{\phi}_{\bmA\bmB\bmC\bmD}$,
$\bar{\Theta}_{\bmA\bmB\bmC\bmD}$ and $\bar{\varphi}_{\bmA\bmB}$.  Under the
rescaling $\delta_\bmA\mapsto \bar{\kappa}^{1/2} \delta_\bmA$, the latter
fields can be see to transform as
\[
\bar{\phi}_{\bmA\bmB\bmC\bmD} \mapsto \bar{\kappa}^3 \bar{\phi}_{\bmA\bmB\bmC\bmD},
\qquad 
\bar{\Theta}_{\bmA\bmB\bmC\bmD} \mapsto \bar{\kappa}^2 \bar{\Theta}_{\bmA\bmB\bmC\bmD},
\qquad
\bar{\varphi}_{\bmA\bmB} \mapsto \bar{\kappa}^2 \bar{\varphi}_{\bmA\bmB}.
\]
One also has that
\[
\bar{\bme}_{\bmA\bmB} \mapsto \bar{\kappa} \bar{\bme}_{\bmA\bmB}, \qquad
\bar{\bmsigma}^{\bmA\bmB} \mapsto \bar{\kappa}^{-1} \bar{\bmsigma}^{\bmA\bmB}, \qquad
\bar{\gamma}_{\bmA\bmB\bmC\bmD} \mapsto \bar{\kappa} \gamma_{\bmA\bmB\bmC\bmD}
-\tfrac{1}{2}(\epsilon_{\bmA\bmC} \bar{\bme}_{\bmB\bmD}\bar{\kappa} +
\epsilon_{\bmB\bmD} \bar{\bme}_{\bmA\bmC} \bar{\kappa}).
\]
This is completely analogous to the rescaling behaviour for fields in
$\mathcal{C}_{a,\kappa}$ as discussed in Section
\ref{Section:ExtendedBundleSpace}. A calculation readily shows that
the lifts of the fields to $\bar{\mathcal{C}}_{\bar{a}}$ is given by
\[
\bar{\phi}_{\bmA\bmB\bmC\bmD} = -
\frac{\displaystyle 6m}{r^2(\displaystyle
 r+m)}\epsilon^2_{\bmA\bmB\bmC\bmD}, 
\qquad  \bar{\Theta}_{\bmA\bmB\bmC\bmD} = \frac{\displaystyle 6m}{\displaystyle
  r (r+m)^2}\epsilon^2_{\bmA\bmB\bmC\bmD}, 
\qquad \bar{\varphi}_{\bmA\bmB} = \frac{m}{r^2} x_{\bmA\bmB}.
\]
Thus it follows that one should chose
$\bar{\kappa}=O(r)$ in order to obtain rescaled fields which are
finite at $r=0$. In what follows we make the simplest choice
\begin{equation}
\bar{\kappa} =r.
\label{Choicebarkappa}
\end{equation}
As a result one finds that the lift of the 3-metric $\bar{\bmh}=-\bar{\bmdelta}$ satisfies
\[
\bar{\bmh} \mapsto \frac{1}{r^2} \mathbf{d}r\otimes \mathbf{d}r +
2(\bmalpha^+ \otimes \bmalpha^- + \bmalpha^- \otimes \bmalpha^+).
\]
Hence, as in the case of the cylinder at spatial infinity one obtains
a 3-metric which is singular at $r=0$. Again, this is not a problem
for our general strategy of obtaining a regular representation of the
spacetime region around $i^+$ and $c^0$ as the conformal field
equations are expressed in terms of quantities which are chosen to be
regular at $r=0$. 

\subsection{Spacetime gauge considerations}
The implementation of a spacetime gauge on the domain of dependence of
$\bar{\mathcal{C}}_{\bar{a},\bar{\kappa}}$ is done in complete analogy
to the construction of the cylinder at infinity. Recall that at the
cylinder-like end of $\tilde{\mathcal{S}}$ in region $I$ and
$\tilde{\regionIIIslices}$ in region $\widetilde{III}$ the conformal
factor was chosen to be $\varpi = r/(r+m)$. We prescribe the following
initial data on $\mathcal{B}_a(\bar{\i})\setminus \{ \bar{\i} \}$ for
a congruence of conformal geodesics
\begin{equation}
\bar{\bmx}_\star = \bar{\bmx}(p), \qquad \dot{\bar{\bmx}}_\star \perp
\mathcal{S}_{\bar{\i}}, \qquad
\tilde{\bar{\bmb}}_\star = \varpi^{-1} \mathbf{d} \varpi, \qquad \langle
\tilde{\bmb}, \dot{\bar{\bmx}}
\rangle_\star =0.
\label{ConformalGeodesicsDataHorizon}
\end{equation}
The above initial data is supplemented by the following
choice of data for the conformal factor
\[
\bar{\Theta}_\star = \bar{\kappa}^{-1} \varpi, \qquad \dot{\bar{\Theta}}_\star =0, \qquad
\ddot{\bar{\Theta}}_\star =
-\frac{\bar{\kappa}}{2\varpi}\bar{\bmdelta}(\mathbf{d}\varpi,\mathbf{d}\varpi) +
\bar{\kappa}^{-3} \varpi^3 \bar{\mu}_\star , 
\]
with $\bar{\kappa}$ given as in \eqref{Choicebarkappa}. Hence
$\bar{\Theta}_\star = 1/(r+m)$. In addition we consider a frame $\{
\bar{\bme}_\bma \}$ which is Weyl propagated along the congruence of
conformal geodesics and set $\bar{\bme}_0 =\dot{\bar{\bmx}}$, as
usual. The above initial data conditions ensure that
\[
\bar{\bmg} (\dot{\bar{\bmx}},\dot{\bar{\bmx}}) =\bar{\Theta}^2 \tilde{\bar{\bmg}}
(\dot{\bar{\bmx}},\dot{\bar{\bmx}}) =1
\]
consistently with $\bar{\bmg} = \bar{\Theta}^2 \tilde{\bar{\bmg}}$, 
where $\tilde{\bar{\bmg}}$ denotes
the (physical) spacetime metric associated to the development of
$\mathcal{S}_{\bar{\i}}$. We also require that
\[
\big(\bar{\bmg}(\bar{\bme}_\bma,\bar{\bme}_\bmb)\big)_\star=\bar{\Theta}^2_\star \big(\tilde{\bar{\bmg}} (\bar{\bme}_\bma,\bar{\bme}_\bmb)\big)_\star = \eta_{\bma\bmb}.
\]

\medskip
As in the case of the cylinder at spatial infinity, we consider 3
connections $\tilde{\bar{\bmnabla}}$, $\bar{\bmnabla}$ and $\hat{\bar{\bmnabla}}$
---respectively, the physical Levi-Civita connection, the unphysical
Levi-Civita connection and the canonical Weyl connection associated to
the congruence of conformal geodesics. The relation between these various
connections is given by expressions which are completely analogous to
those described in Section \ref{Section:WeylGaugeConditions} for the cylinder at spatial
infinity. In particular, one has that
\[
\hat{\bar{\bmnabla}} - \bar{\bmnabla} = \bmS(\bar{\bmf}),
\]
where the 1-form $\bar{\bmf}$ satisfies
\[
\bar{\bmf} = \tilde{\bar{\bmb}} - \bar{\Theta}^{-1}
\mathbf{d}\bar{\Theta}. 
\]
By analogy to the gauge construction in the cylinder at spatial
infinity one obtains, along the congruence of conformal geodesics, the
gauge conditions
\[
\hat{\bar{\Gamma}}_0{}^\bmb{}_\bmc =0, \qquad \bar{f}_0 =0, \qquad \hat{\bar{L}}_{0\bma}=0, 
\]
for the Weyl connection coefficients, the time component of
$\bar{\bmf}$ and the components of the Schouten tensor of the Weyl
connection $\hat{\bar{\bmnabla}}$, respectively. 

\subsubsection{A representation of the horizon}
In the sequel, it will be seen that using
the conformal isometry of the Reissner-Nordstr\"om, the  conformal geodesics in the
construction of the cylinder at spatial infinity can be mapped to the
conformal geodesics in a neighbourhood of $\bar{\i}$ arising from the
data \eqref{ConformalGeodesicsDataHorizon}. As a consequence, given
$q\in \mathcal{B}_{\bar{a}}(\bar{\i})$, there exists $a>0$ and $p\in \mathcal{B}_a(i^0)$
such that $q = \iota (p)$. 
Recall that in Section \ref{Section:StructureConformalBoundary},
$\tau_\mathscr{I}(p)$ was defined by the condition
$\Theta(\pm\tau_\mathscr{I}(p))=0$ with $\Theta$ the conformal factor
associated to the congruence of conformal geodesics near the cylinder
at spatial infinity. Let $\tau_{\mathscr{H}} \equiv \iota_*
\tau_\mathscr{I}$.  As a result of the properties of the conformal
isometry discussed in Section \ref{Section:ConformalInversion} it
follows that $\bar{\Theta}(\pm \tau_\mathscr{H}(q))\neq 0$.
Nevertheless, the function $\tau_{\mathscr{H}} \equiv \tau_\mathscr{I}
\circ \iota$ will be useful to identify the domain on which we will be
looking for solutions to the conformal Einstein-Maxwell equations
---namely:
\[
\bar{\mathcal{M}}_{\bar{a},\bar{\kappa}} \equiv \{ (\tau, q) \in \mathbb{R}\times
\bar{\mathcal{C}}_{\bar{a},\bar{\kappa}} \; | \;  -\tau_{\mathscr{H}}(q)\leq \tau \leq \tau_{\mathscr{H}}(q)
\}. 
\]
We also define the \emph{horizon} $\mathscr{H}=\mathscr{H}^+\cup \mathscr{H}^-$ with
\[
\mathscr{H}^\pm \equiv \{  (\tau,q) \in
\bar{\mathcal{M}}_{\bar{a},\bar{\kappa}} \; | \;  \tau =\pm \tau_{\mathscr{H}}(q), \; r(q) \neq 0\},
\]
and the \emph{cylinder at the horizon} as 
\[
\cylinderhorizon \equiv \{ (\tau,q) \in \bar{\mathcal{M}}_{\bar{a},\bar{\kappa}} \; | \;
r(q) =0, \,  -\tau_{\mathscr{H}}(q) < \tau < \tau_{\mathscr{H}}(q) \}.
\]
Of interest are also the \emph{critical sets}
\[
\cylinderhorizon^\pm \equiv \{ (\tau,q) \in \bar{\mathcal{M}}_{\bar{a},\bar{\kappa}} \; | \;
r(q) =0, \,  \tau =\pm \tau_{\mathscr{H}}(q) \},
\]
and
\[
\cylinderhorizon^0 \equiv \{ (\tau,q) \in \bar{\mathcal{M}}_{\bar{a},\bar{\kappa}} \; | \;
r(q) =0, \,  \tau =0 \},
\]
the intersection of $\bar{\mathcal{C}}_{\bar{a},\bar{\kappa}}$ with
$\cylinderhorizon$. 

\medskip
Coordinates in $\bar{\mathcal{M}}_{\bar{a},\bar{\kappa}}$ are
naturally dragged from $\bar{\mathcal{C}}_{\bar{a},\bar{\kappa}}$
along the conformal geodesics. We emphasise that as a result the $r$
coordinate used below is constant along conformal geodesics and hence
does no longer represent the isotropic coordinate. Hence statements
involving $r=0$ refer to the cylinder at the horizon
$\cylinderhorizon$, not the horizon $\mathscr{H}^\pm $. The former is
ruled by the conformal geodesics starting at $r=0$.  Similarly, the
vector fields $\bmpartial_r$, $\bmX_\pm$ extend in a unique way to
vectors on $\bar{\mathcal{M}}_{\bar{a},\bar{\kappa}}$ by requiring
that they commute with $\bmpartial_\tau$.

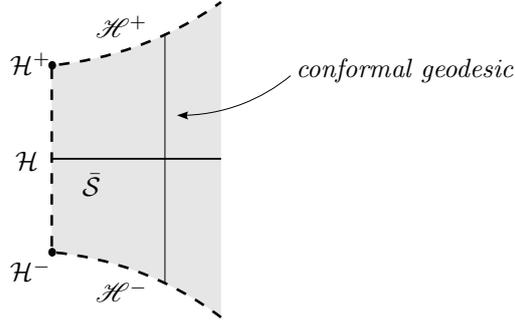
\begin{figure}[t]
\begin{center}
\includegraphics[width=0.25\textwidth]{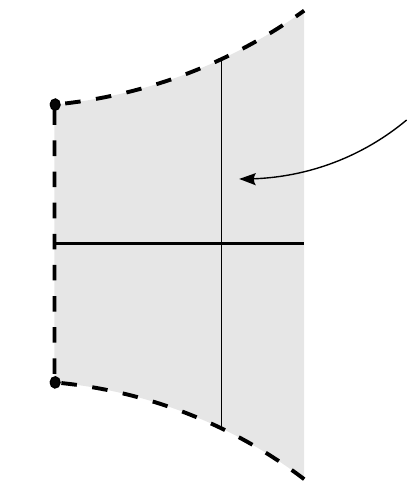}
\put(-75,110){$\mathscr{H}^+$}
\put(-75,10){$\mathscr{H}^-$}
\put(-80,50){$\bar{\mathcal{S}}$}
\put(-107,96){$\cylinderhorizon^+$}
\put(-107,18){$\cylinderhorizon^-$}
\put(-105,60){$\cylinderhorizon$}
\put(0,95){\emph{conformal geodesic}}
\end{center}
\caption{Schematic 2-dimensional diagram of the bundle manifold
  $\bar{\mathcal{M}}_{\bar{a},\bar{\kappa}}$ giving rise to the cylinders at the
  horizon. The null hypersurfaces $\mathscr{H}^\pm$ correspond to the
  horizon, $\cylinderhorizon$ is the cylinder at the horizon proper
  and $\cylinderhorizon^\pm$ are the critical sets where the cylinder
meets the horizon. The conformal geodesics on which this construction
is based correspond to straight vertical lines as indicated. Notice
that the diagram is not conformal ---null hypersurfaces are not
straight lines with slope of  $45^\circ$ or $135^\circ$. The boundary
of  $\bar{\mathcal{M}}_{\bar{a},\bar{\kappa}}$ has been depicted with a
broken line to indicate that it does not belong to conformal boundary.} 
\label{Figure:CylinderHorizon}
\end{figure}

\subsection{The regular initial value problems at $c^0$ and $i^+$}
\label{Section:RegularInitialValueProblemHorizon}

A regular initial value problem for the conformal field equations
\eqref{SpinorialXCFE} with data on the bundle manifold
$\bar{\mathcal{C}}_{\bar{a},\bar{\kappa}}$ can be posed in analogy to that
for the cylinder at spatial infinity discussed in Sections
\ref{Section:SSEvolutionEquations} and \ref{Section:InitialData}. 
Analogous to Section \ref{Section:SSEvolutionEquations} one makes
use of a spherically symmetric Ansatz and one obtains ordinary
differential equations along the curves of the congruence of conformal
geodesics for the fields
\[
\bar{\mathbf{u}} \equiv \big( \bar{e}^0, \, \bar{e}^1, \bar{e}^\pm, \bar{f}, \bar{\xi}_x, \bar{\chi}_2 , \bar{\chi}_h,
\bar{\theta}_x, \bar{\theta}_2, \bar{\theta}_h, \bar{\phi}, \bar{\varphi}, \bar{\Theta}, \bar{\Sigma}, \bar{\Lambda}\big).
\]
 The regular data
 for these equations is given by 
\begin{subequations}
\begin{eqnarray}
& \bar{e}^0=0, \quad \bar{e}^1=r, \quad \bar{e}^\pm=1,& \label{GeometricDataHorizon1}\\
& \bar{f}=1, \quad \bar{\xi}_x =0, \quad \bar{\chi}_2=0, \quad \bar{\chi}_h=0& \\
& \bar{\theta}_2=\frac{\displaystyle 6m}{\displaystyle (r+m)^2}, \quad \bar{\theta}_2 =0, \quad \bar{\theta}_h =0, \quad \bar{\phi}=-\frac{\displaystyle 6m}{\displaystyle r+m}, \quad \bar{\varphi}=m .& \label{GeometricDataHorizon2}
\end{eqnarray}
\end{subequations}
The initial data for the conformal gauge unknowns is given by
\begin{subequations}
\begin{eqnarray}
& \bar{\Theta}_\star = \frac{\displaystyle 1}{\displaystyle r+m}, \qquad \dot{\bar{\Theta}}_\star =0, \qquad  \ddot{\bar{\Theta}}_\star =-\frac{\displaystyle (r+2m)^2}{\displaystyle (r+m)^3} , & \label{GaugeDataHorizon1}\\
& d_{x\star} = \frac{\displaystyle m}{\displaystyle r(r+m)^2}. & \label{dxDataHorizon}
\end{eqnarray}
\end{subequations}
In particular notice that $\bar{\Theta}_\star,
\;\ddot{\bar{\Theta}}_\star \neq 0$ at $r=0$.

\medskip
The equations governing the evolution of the components of $\bar{\mathbf{u}}$ are \emph{formally identical} to equations
\eqref{Reduced:SSEe0}-\eqref{Reduced:SSEvarphi},
\eqref{Reduced:SSETheta} and \eqref{Reduced:SSEd0}. We write this
system of equations and its initial data schematically as
\begin{equation}
\partial_\tau \bar{\mathbf{u}} = \bar{F}(\bar{\mathbf{u}},\tau,r;m), \qquad  \bar{\mathbf{u}}(0,r;m)=\bar{\mathbf{u}}_\star(r;m)
\label{SystemHorizon}
\end{equation}
with $\bar{F}$ and $\bar{\mathbf{u}}_\star(r;m)$ analytic functions of
its arguments. Also, observe that if $m=0$, the initial data for the
conformal factor is singular at $r=0$. Thus, the behaviour of
solutions to the system of transport equations \eqref{SystemHorizon}
cannot be analysed directly by means of a perturbative argument as in
the case of the cylinder at spatial infinity. Given a solution to system \eqref{SystemHorizon} one readily can construct an unphysical metric $\bar{\bmg}$ by analogy to \eqref{SpacetimeMetric} via
\begin{equation}
\bar{\bmg} =\mathbf{d}\tau \otimes \mathbf{d}\tau -
\frac{\bar{e}^0_x}{\bar{e}^1_x}\left( \mathbf{d}\tau \otimes \mathbf{d}\rho +
  \mathbf{d}\rho \otimes \mathbf{d}\tau \right) - \left(
  \frac{1}{(\bar{e}^1_x)^2} -\left( \frac{\bar{e}^0_x}{\bar{e}^1_x}\right)^2 \right)
\mathbf{d}\rho \otimes \mathbf{d}\rho -\frac{1}{\bar{e}^+_z
  \bar{e}^-_y}\bmsigma.
\label{UnphysicalMetricHorizon}
\end{equation}

\medskip
In order to obtain a suitable existence result which exhausts the whole of
$\bar{\mathcal{M}}_{\bar{a},\bar{\kappa}}$ for sufficiently small $\bar{a}>0$, we
exploit the conformal isometry $\iota$ to map the solution
$\mathbf{u}$ of system \eqref{EvolutionSystem} on $\mathcal{M}_{a,\kappa}$ to
the required solution $\bar{\mathbf{u}}$ of system \eqref{SystemHorizon}. One obtains the following result:

\begin{theorem}
\label{Theorem:Main}
Given $m>0$, there exists $\bar{a}>0$ such that there exists a unique
smooth solution to
the spherically symmetric regular initial value problem at $c^0$
(or $i^+$) for the extended conformal Einstein-Maxwell equations on the
whole of $\bar{\mathcal{M}}_{\bar{a},\bar{\kappa}}$. The
surfaces $\mathscr{H}^\pm$ corresponds to the horizons of the physical
metric $\tilde{\bar{\bmg}}=\bar{\Theta}^{-2} \bar{\bmg}$ with
$\bar{\bmg}$ given by equation \eqref{UnphysicalMetricHorizon}.
\end{theorem}

The proof of this, our main result, follows from the discussion of the following sections.

\subsection{The action of $\iota$ on $\mathcal{M}_{a,\kappa}$ and $\bar{\mathcal{M}}_{\bar{a},\bar{\kappa}}$}
In view that the construction of the manifolds
$\mathcal{M}_{a,\kappa}$ and
$\bar{\mathcal{M}}_{\bar{a},\bar{\kappa}}$ is based on the properties
of conformal invariants, it is natural to expect a nice transformation
between the objects on one manifold and the other under the action of
the conformal isometry $\iota$. In this section we discuss these
transformations. For conciseness some of the details are put in an
appendix at the end of the article. Recall that quantities near the
horizon are distinguished from those near $i^0$ by an overbar and
physical quantities are denoted with a tilde over them.  One can think
of working with two identical manifolds $(\bar{M}, \tilde{\bar{g}})$
and $(M,\tilde{g})$, where the points are interpreted differently. The
horizon in one copy corresponds to the conformal boundary in the
other, and so. Under the conformal isometry $\iota_* \tilde{\bmg}
=\Omega^2 \tilde{\bar{\bmg}}$.

\subsubsection{The correspondence between the congruence of conformal geodesics}
As already discussed in Section \ref{Section:CGIsometry}, the conformal isometry maps conformal geodesics into conformal geodesics. More, precisely, as a consequence of formula \eqref{GCIsometryAction}, conformal geodesics with initial data 
\begin{eqnarray*}
& p \in \tilde{\mathcal{S}}, \qquad \bmx_\star= \bmx(p), \qquad \dot{\bmx}_\star\perp \tilde{\mathcal{S}},& \\
&\tilde{\bmb}_\star = \omega^{-1} \mathbf{d} \omega &
\end{eqnarray*}
are mapped into conformal geodesics satisfying
\begin{eqnarray*}
&q=\iota(p) \in \tilde{\bar{\mathcal{S}}},\quad
\bar{\bmx}_\star=\bar{\bmx}(q),  \quad \dot{\bar{\bmx}}_\star \perp
\tilde{\bar{\mathcal{S}}}, & \\
&\tilde{\bar{\bmb}}_\star = \iota_*(\omega^{-1} \mathbf{d} \omega) + \Omega^{-1} \mathbf{d}\Omega = \varpi^{-1} \mathbf{d}\varpi.&
\end{eqnarray*}
This can be verified by an explicit computation noticing that
 \[
\omega^{-1} \mathbf{d}\omega =
\frac{2+m\rho}{\rho(1+m\rho)}\mathbf{d}\rho =
 \frac{\varrho +  2m}{\varrho (\varrho +m )} \mathbf{d}\varrho, 
 \qquad \iota_* (\omega^{-1}\mathbf{d}\omega) = 
 \frac{r +  2m}{r (r +m )} \mathbf{d}r, 
 \qquad \Omega^{-1}\mathbf{d}\Omega = -\frac{1}{r} \mathbf{d}r,
\]
and that
\[
\varpi^{-1}\mathbf{d}\varpi = \frac{m}{r(r+m)} \mathbf{d}r.
\]

\medskip
As a consequence of the above discussion, the curves $\bmx(\tau)$ and
$\bar{\bmx}(\tau)$ with initial data as given above share the same
affine parameter. Moreover, the radial coordinates $\rho$ and $r$ on
the conformal extensions $\mathcal{S}_{i^0}$ and
$\mathcal{S}_{\bar{\i}}$ are related to each other by
\[
\rho \stackrel{\iota}{\longmapsto} r/m^2.
\]
Thus, the conformal isometry $\iota$ induces a correspondence between
the conformal Gaussian coordinates of $\mathcal{M}_{a,\kappa}$ and
$\bar{\mathcal{M}}_{\bar{a},\bar{\kappa}}$ which is given by
\[
(\tau,\rho, t^\bmA{}_\bmB) \stackrel{\iota}{\longmapsto} (\tau,
r/m^2,t^\bmA{}_\bmB), \qquad t^\bmA{}_\bmB \in SU(2,\mathbb{C}).
\]
Accordingly, one has that
\[
\iota (\mathscr{I})^\pm = \mathscr{H}^\pm, \qquad \iota(\cylinderinfinity) =
\cylinderhorizon, \qquad \iota(\cylinderinfinity^\pm) = \cylinderhorizon^\pm,
\]
and one has a natural bijection between $\mathcal{M}_{a,\kappa}$ and
$\bar{\mathcal{M}}_{\bar{a},\bar{\kappa}}$ with $\bar{a}=a/m^2$. In what follows, it will be seen
that this bijection is, in fact, an isometry.

\subsubsection{The correspondence between the conformal factors}

In order to relate the conformal factors $\Theta$ and $\bar{\Theta}$
leading, respectively, to the representations of the cylinder at
spatial infinity and the cylinders at the horizon, one starts by
recalling that these are defined by the conditions
\begin{equation}
 \Theta^2\tilde{\bmg}(\dot{\bmx},\dot{\bmx})=1, \qquad \bar{\Theta}^2 \tilde{\bar{\bmg}}(\dot{\bar{\bmx}},\dot{\bar{\bmx}})=1.
\label{DefinitionCanonicalConformalFactors}
\end{equation}
Applying the inversion $\iota$ to the first of these relations, and
recalling that $\iota_* \tilde{\bmg} =\Omega^2 \tilde{\bar{\bmg}}$, one has
that
\begin{eqnarray*}
&& \big(\Theta^2  \tilde{\bmg}(\dot{\bmx},\dot{\bmx})
\big)\circ \iota = (\Theta\circ \iota)^2 \iota_*
\tilde{\bar{\bmg}}(\iota_* \dot{\bmx},\iota_*\dot{\bmx}) \\
&& \phantom{\big(\Theta^2  \tilde{\bmg}(\dot{\bmx},\dot{\bmx})
\big)\circ \iota} = (\Theta\circ \iota)^2 \Omega^2 \tilde{\bar{\bmg}}(\dot{\bar{\bmx}},\dot{\bar{\bmx}})=1.
\end{eqnarray*}
Comparing with the second relation in
\eqref{DefinitionCanonicalConformalFactors} one concludes that
\begin{equation}
\bar{\Theta} = (\Theta\circ \iota) \Omega. 
\label{ThetaRelation}
\end{equation}
In order to find a relation between the conformal metrics
$\bmg$ and $\bar{\bmg}$ recall that
\begin{equation}
\bmg= \Theta^2 \tilde{\bmg}, \qquad \bar{\bmg} = \bar{\Theta}^2 \tilde{\bar{\bmg}}.
\label{DefinitionConformalMetrics}
\end{equation}
Writing the first of the relations in
\eqref{DefinitionConformalMetrics} in the form $\tilde{\bmg}
=\Theta^{-2}\bmg$ one has that
\[
\iota_* \tilde{\bmg}= \iota_* \big( \Theta^{-2} \bmg \big)
= \big(\Theta\circ \iota)^{-2} \iota_* \bmg.
\]
Recalling that $\iota_* \tilde{\bmg} =\Omega^2 \tilde{\bar{\bmg}}$ one
concludes that
\[
\iota_* \bmg= \big( \Theta \circ \iota \big)^2 \Omega^2 \tilde{\bar{\bmg}}.
\]
From the second relation in \eqref{DefinitionConformalMetrics}
in the form $\tilde{\bar{\bmg}}= \bar{\Theta}^{-2} \bar{\bmg}$ one finds that
\[
\iota_* \bmg = \big(\Theta\circ \iota \big)^2 \Omega^2 \bar{\Theta}^{-2} \bar{\bmg}.
\]
Hence, using \eqref{ThetaRelation} one concludes that
\begin{equation}
\iota_* \bmg =\bar{\bmg} \quad \mathrm{and} \quad \iota_* \bar{\bmg} = \bmg
\label{RelationConformalMetrics}
\end{equation}
\emph{In other words, the conformal metrics $g$ and $\bar{g}$ leading to the cylinder representations
are related to each other by an isometry}. In particular, if $\{
\bme \}$ denotes a $\bmg$-orthogonal frame, then $\{ \bar{\bme}_\bma
\} \equiv \{ \iota_* \bme_\bma \}$ is a $\bar{\bmg}$ orthogonal
frame. If $\{ \omega_\bma \}$ is the associated coframe of
$\{\bme_\bma\}$, then $\{ \bmomega^\bma\} \equiv \{\iota_*
\bar{\bmomega}^\bma \}$ is the associated coframe of
$\{\bar{\bme}_\bma\}$  ---i.e. $\langle \bar{\bmomega}^\bmb,
\bar{\bme}_\bma \rangle = \delta_\bma{}^\bmb$.

\subsubsection{Correspondence between the various conformal fields}
\label{Section:TransformationFormulaeInversionConformalFields}

In order to establish Theorem \ref{Theorem:Main} one needs to examine
the explicit correspondence between the conformal fields appearing in
the conformal field equations on $\mathcal{M}_{a,\kappa}$ and those
appearing on the equations on $\bar{\mathcal{M}}_{\bar{a},\bar{\kappa}}$.

\medskip
Explicit computations show that
\begin{eqnarray*}
& \bar{f}_\bma = \iota_* f_\bma, \qquad
\bar{\Gamma}_\bma{}^\bmb{}_\bmc = \iota_* \Gamma_\bma{}^\bmb{}_\bmc, & \\
&  \bar{d}^\bma{}_{\bmb\bmc\bmd}= \Omega^{-1} \,
\iota_* d^\bma{}_{\bmb\bmc\bmd}, \qquad \hat{\bar{L}}_{\bma\bmb} =
\iota_* \hat{L}_{\bma\bmb} \qquad \bar{E}_\bma = \iota_* E_\bma& \\
& \bar{\eta}_\bma = \iota_* \eta_\bma& 
\end{eqnarray*}
where $\iota_* f_\bma\equiv f_\bma\circ \iota$, etc. 
The details are given in the appendix.
From these frame
expressions one readily obtains the transformation for the spinorial
components. These are given by
\begin{eqnarray*}
&\bar{f} = \iota_* f, \quad \bar{\xi}_x = \iota_* \bar{\xi}_x, \quad
\bar{\chi}_2 = \iota_* \chi_2, \quad \bar{\chi}_h = \iota_* \chi_h & \\
& \bar{\phi}= \Omega^{-1} \iota_* \phi, \quad \bar{\theta}_x = \iota_*
\theta_x, \quad \bar{\theta}_2 = \iota_* \theta_2, \quad
\bar{\theta}_h =\iota_* \theta_h, \quad \bar{\varphi}= \iota_* \varphi.
\end{eqnarray*}

\medskip
\noindent
\textbf{Remark.} On the (unphysical) conformal fields $\iota$ acts as
an isometry, while for the physical variables $\iota$ acts as a
conformal isometry.

\medskip
As the fields
\[
f, \quad \xi_x,\quad  \bar{\chi}_2, \quad 
\bar{\chi}_h, \quad  \bar{\theta}_x, \quad  \bar{\theta}_2, \quad
\bar{\theta}_h, \quad \bar{\varphi}
\] 
are regular on $\mathcal{M}_{a,\kappa}$, it follows directly from the above
transformation rules that
\[
\bar{f}, \quad \bar{\xi}_x,\quad \bar{\chi}_2, \quad
\bar{\chi}_h,\quad  \bar{\theta}_x, \quad  \bar{\theta}_2,\quad
\bar{\theta}_h, \quad \bar{\varphi}
\]
are also regular on the whole of
$\bar{\mathcal{M}}_{\bar{a},\bar{\kappa}}$. Moreover, it follows
that $\bar{\eta}_x \neq 0$ on
$\bar{\mathcal{M}}_{\bar{a},\bar{\kappa}}$ for suitably small $a>0$,
so no conjugate points arise in the congruence of conformal geodesics.

\medskip
In order to conclude our analysis it only remains to consider the
behaviour of $\bar{\phi}$ and $\bar{\Theta}$ at the horizon. This analysis requires information
about the behaviour of the solutions at $\cylinderhorizon$.

\subsection{The solution to the conformal field equations near the cylinders at the horizon}
The behaviour of the solutions of the regular initial value problem formulated in Section
\ref{Section:RegularInitialValueProblemHorizon} on the cylinder
$\cylinderhorizon$ is of particular interest.

\medskip
Combining the expressions in \eqref{SolutionCylinderSpatialInfinity}
for the solutions of the regular initial value problem at spatial
infinity with the transformation formulae discussed in the previous
section one readily finds that
\[
\bar{\xi}_2= \bar{\xi}_h = \bar{\chi}_2 =\bar{\chi}_h = \bar{\theta}_x=\bar{\theta}_2 =\bar{\theta}_h =0 \qquad \mbox{on } \cylinderhorizon.
\]
Under these conditions the evolution equations for the conformal
factor and the components of the Weyl and Maxwell spinors reduce to:
\begin{subequations}
\begin{eqnarray}
&& \dot{\bar{\phi}} = \dot{\bar{\Theta}} \bar{\varphi}^2, \label{CylinderEquation1}\\
&& \dot{\bar{\varphi}}=0, \label{CylinderEquation2} \\
&& \dddot{\bar{\Theta}} = 2\bar{\Theta}^2 \dot{\bar{\Theta}} \bar{\varphi}^2. \label{CylinderEquation3}
\end{eqnarray}
\end{subequations} 
Using the constraint equations for $\dot{\bar{\Theta}}_\star$ and
$\ddot{\bar{\Theta}}_\star$ on finds that the initial data for the above
equations is given by
\[
\bar{\varphi}_\star= -m, \qquad \bar{\phi}_\star=0,
\qquad \bar{\Theta}_\star = \frac{1}{m}, \qquad
\dot{\bar{\Theta}}_\star =0, \qquad \ddot{\bar{\Theta}}_\star
=0, \qquad \mbox{on } \cylinderhorizon^0. 
\]
It follows that the solution to the above equations is constant along
$\cylinderhorizon$. Accordingly, one has that
\[
\bar{\varphi}=-m, \qquad \bar{\phi} =0, \qquad
\bar{\Theta} = \frac{1}{m}, \quad \mbox{on} \quad \cylinderhorizon.
\]
Notice, in particular that the conformal factor $\bar{\Theta}$ does
not vanish on the cylinder $\cylinderhorizon$. Thus,
$\cylinderhorizon$ is not part of the conformal boundary. Moreover, by
continuity, it follows that at least suitably close to
$\cylinderhorizon$, $\bar{\Theta}\neq 0$ on $\mathscr{H}^+$.

\subsubsection{Behaviour of $\bar{\phi}$ at the horizon}
Finally, to conclude the analysis, we consider the behaviour of
$\bar{\phi}$ on $\mathscr{H}^+$. To this end recall the transformation
formula $\bar{\phi}= \Omega^{-1} \iota_* \phi$. To show that $\bar{\phi}$ is regular on $\mathscr{H}^+$
consider the analogue of the \emph{core system equation} \eqref{Reduced:X} on
$\bar{\mathcal{M}}_{\bar{a},\bar{\kappa}}$:
\[
\partial_\tau \bar{X}=-\tfrac{1}{3} \bar{X}^2 -\bar{L} + \tfrac{1}{2} \bar{\Theta} \bar{\phi}.
\]
As $\bar{X}$, $\bar{L}$ and $\bar{\Theta}$ are regular on
$\mathscr{H}^+$, and moreover $\bar{\Theta}\neq 0$ on $\mathscr{H}^+$
at least sufficiently close to $\cylinderhorizon$, it follows that
$\bar{\phi}$ must be regular at $\mathscr{H}^+$, at least close enough
to $\cylinderhorizon$. In fact, one can say a bit more by noticing
that $\Omega^{-1} =r/m$ so that $\Omega^{-1}=0$ at the horizon.  As
$\phi$ is regular on $\mathscr{I}^+$, it follows that $\iota_* \phi$ is
regular at $\mathscr{H}^+$ and $\bar{\phi}$ vanishes at the horizon. 

\medskip
With this argument we conclude the proof of Theorem \ref{Theorem:Main}.

\section{Concluding remarks and perpectives}
\label{Section:ConcludingRemarks}
The main conclusion of our analysis is
that it is possible to obtain regular conformal representations of the
extremal Reissner-Nordstr\"om spacetime in a neighbourhood of the
points $i^+$ and $c^0$. These conformal representations have been
obtained as the solution of the conformal field equations written in
terms of a gauge based on conformal invariants of the spacetimes
(conformal geodesics). It is important to point out that this
representation is regular for the variables appearing in the conformal
field equations, but not for the associated conformal (unphysical)
metric. This peculiarity is not a major source of problems as the
metric is not one of the field unknowns one is solving for.

Our construction of the cylinders at the horizon and the subsequent analysis have been eased by the conformal discrete isometry in the extremal Reissner-Nordstr\"om inducing an isometry in the unphysical setting. However, we claim that the
existence of this (conformal) isometry is not essential for the analysis. It is
just a convenient property to shorten some of the arguments and to
gain insight into the underlying structures. In particular, we claim it
should be possible to obtain an analogous representation for the
extremal Kerr spacetime.

In the remainder of this section we discuss some possible implications of the
present construction of the cylinders at the horizon. 

\subsubsection*{Behaviour of test fields near the cylinders at the horizon}
The key insight obtained from the construction of the cylinder at
spatial infinity for vacuum spacetimes given in \cite{Fri98a} is the
existence of logarithmic singularities at
the critical sets $\cylinderinfinity^\pm$ for generic initial data. 
Given that null infinity is a characteristic of the field equations, 
it is expected that these singularities will spread along $\mathscr{I}$, thus giving rise to conformal boundary which is
non-smooth. Although this picture seems quite plausible, there is no proof
available for this conjecture. There is, however, some analysis with
linear test fields which suggests how the full non-linear case could be
controlled ---see \cite{Fri03b,Val09a}.

Although an analysis like the one described in the previous paragraph
has not be carried out for electrovacuum spacetimes, it is reasonable
to expect a similar behaviour at the critical sets of the cylinder at
spatial infinity. More precisely, given  the spin-1
zero-rest mass field equation
\begin{equation}
\nabla^{AA'} \lambda_{AB} =0, \qquad \lambda_{ABC}
=\lambda_{(AB)}. 
\label{Spin1Equation}
\end{equation}
a standard hyperbolic reduction procedure leads to an evolution
system of the form 
\begin{equation}
(\sqrt{2} \mathbf{E} + \mathbf{A}^{\bmA\bmB} e^0_{\bmA\bmB})\partial_\tau \bmlambda + \mathbf{A}^{\bmA\bmB}
e^\alpha_{\bmA\bmB} \partial_\alpha \bmlambda = \mathbf{B}(\bmGamma)\bmlambda.
\label{SHSTestField}
\end{equation}
Here $\bmlambda$ is a column vector with complex valued components
$\lambda_0, \, \lambda_1,\, \lambda_2$ (the independent components of
$\lambda_{AB}$), and where $\mathbf{A}^{\bmA\bmB}$ are constant
matrices, $\mathbf{B}(\bmGamma)$ denotes a linear matrix value function
of the connection coefficients and $\mathbf{E}$ is the $3\times 3$
unit matrix.  Using the information about the extremal
Reissner-Nordstr\"om spacetime on the cylinder at spatial infinity one readily
finds that
\[
(\sqrt{2} \mathbf{E} + \mathbf{A}^{\bmA\bmB} e^0_{\bmA\bmB}) =
\sqrt{2}
\left(
\begin{array}{ccc}
1+\tau & 0 & 0 \\
0 & 1 & 0 \\
0 & 0 & 1-\tau
\end{array}
\right)
\quad \mbox{ on } \quad \cylinderhorizon.
\]
Thus, the symmetric hyperbolic system \eqref{SHSTestField} degenerates
at the points for which $\tau=\pm 1$. It is likely that this
degeneracy will give rise to logarithmic singularities in the
components of $\bmlambda$ as the solutions approach
$\cylinderhorizon^\pm$ along $\cylinderhorizon$. This feature is expected to appear
in both the cylinders at $i^+$ and $c^0$. The effects of this
degeneracy in the evolution equations and the associated 
singular behaviour will be analysed in detail elsewhere. 

\medskip
If, as anticipated, the correspondence between the cylinders $\cylinderinfinity$ at spatial
infinity and the cylinders $\cylinderhorizon $ at the horizon holds fully, it is to be
expected that these singularities will give rise \emph{polyhomogeneous
  behaviour} of the test fields at the horizon. This potential
non-smoothness of generic test fields at the horizon may be related to
the existence of conserved modes at the horizon in solutions of the wave equation on
the extremal Reissner-Nordstr\"om spacetime \cite{Are11b,Are11a}.

A more intriguing possibility is the idea of performing an analysis of
the full conformal Einstein equations at the cylinders at the horizon
in the manner of \cite{Fri98a} for an initial data set which is a
perturbation of data for the extremal Reissner-Nordstr\"om
spacetime. Although purely asymptotic, this analysis would involve the
full nonlinearities of the Einstein field equations. Consequently, it should
provide valuable insights and have implications for the
question of the non-linear stability of the extremal
Reissner-Nordstr\"om spacetime.

\section*{Acknowledgements}
We have been benefited from
discussions with H. Friedrich and S. Dain on various aspects of the
extremal Reissner-Nordstr\"om spacetime.  JAVK would like to acknowledge the hospitality of the Institute of Mathematics and Applied Systems of the Chinese Academy of Sciences,
Beijing, for its hospitality during a visit in which the last stages
of this project have been carried out. CL would like to acknowledge financial support by the grant
CERN/FP/123609/2011 of FCT and CERN.

\appendix

\section{Transformation formulae under the conformal inversion}

The purpose of the present appendix is to provide some further details
about the derivation of the transformation formulae of various
conformal fields under the conformal isometry $\iota$ given in Section
\ref{Section:TransformationFormulaeInversionConformalFields}. The key
observation in these computations is the isometry $\iota_* \bmg =
\bar{\bmg}$.

\subsubsection*{Transformation of the Weyl tensor}
Exploiting the conformal invariance of the Weyl tensor one has that
\[
C^a{}_{bcd} [\bar{\bmg}] = C^a{}_{bcd} [\bar{\Theta}^2 \tilde{\bar{\bmg}}]= C^a{}_{bcd}[\iota_*
\tilde{\bmg}]  = \iota_* C^a{}_{bcd}[\tilde{\bmg}]= \iota_* C^a{}_{bcd}[\bmg],
\]
where the last equality follows from the fact that $\iota$ is an
isometry. Of more interest for the conformal field equations is the rescaled
Weyl tensor $d^a{}_{bcd}$. Proceeding as in the case of the standard
Weyl tensor one has that
\begin{eqnarray*}
&& d^a{}_{bcd} [\bar{\bmg}] = d^a{}_{bcd} [\bar{\Theta}^2\tilde{\bar{\bmg}}] =\bar{\Theta}^{-1} C^a{}_{bcd}[\Theta^2
\tilde{\bar{\bmg}}] = \bar{\Theta}^{-1} \iota_* C^a{}_{bcd}[\bmg]\\
&& \phantom{d^a{}_{bcd} [\bar{\bmg}]} = \Omega^{-1} \big(
\Theta\circ \iota \big)^{-1} \iota_*
C^a{}_{bcd}[\bmg]= \Omega^{-1} \iota_* \big(
\Theta^{-1} C^a{}_{bcd}[\bmg] \big) = \Omega^{-1} \iota_* d^a{}_{bcd}
[\Theta^2 \tilde{\bmg}],
\end{eqnarray*}
so that one has $d^a{}_{bcd}[\bar{\bmg}] = \Omega^{-1} \iota_*
d^a{}_{bcd}[\bmg]$. From the above expressions one can compute the transformation of
$d^\bma{}_{\bmb\bmc\bmd}$, the components of $d^a{}_{bcd}$ with
respect to the frame $\{\bme_\bma\}$. One finds that
\begin{eqnarray*}
&& \bar{d}^\bma{}_{\bmb\bmc\bmd}\equiv \bar{\bmomega}^\bma{}_a \bar{\bme}_\bmb{}^b
\bar{\bme_\bmc}{}^c \bar{\bme}_\bmd{}^d \bar{d}^a{}_{bcd} = \Omega^{-1} \iota_*
\omega^\bma{}_a \iota_* \bme_\bmb{}^b \iota_*
\bme_\bmc{}^c \bme_\bmd{}^d \iota_* d^a{}_{bcd}\\
&& \phantom{\bar{d}^\bma{}_{\bmb\bmc\bmd}} = \Omega^{-1} (
d^\bma{}_{\bmb\bmc\bmd}\circ \iota) = \Omega^{-1} \iota_*d^\bma{}_{\bmb\bmc\bmd} .
\end{eqnarray*}
The transformation rule for the spinorial components follows
directly from these expressions.

\subsubsection*{Transformation law for the electric field}
In order to analyse that transformation law of the electric field, it
is recalled that the Faraday tensor is  conformally invariant ---that
is, one has that $\tilde{F}_{ab} = F_{ab}$. As a consequence of this
invariance one has that
\begin{eqnarray*}
&& \bar{E}_a \equiv E_a[\bar{\bmg}] = F_{ab}[\bar{\bmg}]
\dot{\bar{x}}^b =F_{ab}[\Theta^2 \tilde{\bmg}] \dot{\bar{x}} ^b = F_{ab}[\iota_*\tilde{\bmg}] \dot{\bar{x}}
^b\\
&&\phantom{\bar{E}_a}= \iota_* F_{ab}[\tilde{\bmg}] \dot{\bar{x}}
^b= \iota_*\big( F_{ab}[\bmg] \dot{x}^b
\big) = \iota_* E_a[\bmg] =\iota_* E_a. 
\end{eqnarray*} 
A similar expression holds for the components with respect to $\{
\bme_\bma \}$ and its spinorial counterpart:
\[
\bar{E}_\bma = \iota_* E_\bma, \qquad \bar{\varphi}_{\bmA\bmB} = \iota_* \varphi_{\bmA\bmB},
\]
so that, in particular, one has that $\bar{\varphi} = \varphi\circ
\iota =\iota_* \varphi$.

\subsubsection*{Transformation law for the connection}
From the discussion in the main text one has that
\[
\bmf = \tilde{\bmb} - \Theta^{-1} \mathbf{d}\Theta, \qquad \bar{\bmf} = \tilde{\bar{\bmb}} - \bar{\Theta}^{-1} \mathbf{d}\bar{\Theta}. 
\]
Hence, it readily follows that
\[
\bar{\bmf} = \iota_* \tilde{\bmb} - ({\Theta}\circ \iota)^{-1}
\mathbf{d}({\Theta}\circ \iota),
\]
so that
\[
\bar{\bmf}=\iota_* {\bmf}. 
\]
Hence, in particular, if $\langle \bmf,
\dot{\bmx}\rangle=0$ then $\langle \bar{\bmf},
\dot{\bar{\bmx}}\rangle=0$. To compute the transformation of the connection coefficients recall
that
\[
\Gamma_\bma{}^\bmb{}_\bmc \equiv \langle \bmomega_\bma, \nabla_\bma
\bme_\bmc \rangle, \qquad \bar{\Gamma}_\bma{}^\bmb{}_\bmc \equiv \langle \bar{\bmomega}_\bma, \bar{\nabla}_\bma
\bar{\bme}_\bmc \rangle.
\]
Hence, using the results from the previous subsections it readily
follows that
\[
\bar{\Gamma}_\bma{}^\bmb{}_\bmc = \iota_* \Gamma_\bma{}^\bmb{}_\bmc =\Gamma_\bma{}^\bmb{}_\bmc\circ \iota.
\]
The spinorial components of $\Gamma_\bma{}^\bmb{}_\bmc$ transform
accordingly. 

\subsubsection*{Transformation law for the Schouten tensor}
The transformation rule of the Schouten tensor follows a similar
procedure as for the other geometric objects. One has that
\[
L_{ab}[\bar{\bmg}] = L_{ab}[\iota_* \bmg]= \iota_* L_{ab}[\bmg].
\]
We write the above more concisely as $\bar{L}_{ab} = \iota_* L_{ab}$.
Moreover, for the Schouten tensors of the corresponding Weyl
connections one notices that
\begin{eqnarray*}
&& \hat{\bar{L}}_{ab} = \bar{L}_{ab} + \bar{\nabla}_a \bar{f}_b - \tfrac{1}{2} \bar{S}_{ab}{}^{cd}
\bar{f}_c \bar{f}_d, \\
&& \phantom{\hat{\bar{L}}_{ab}} = \iota_* L_{ab} + \iota_*
(\nabla_a f_b) - \tfrac{1}{2} \iota_*S_{ab}{}^{cd}
\iota_* f_c \iota_* f_d = \iota_* \hat{L}_{ab}. 
\end{eqnarray*}
In particular for the components with respect to the Weyl propagated
frame one has that $\hat{\bar{L}}_{\bma\bmb} = \iota_* \hat{L}_{\bma\bmb} =
\hat{L}_{\bma\bmb} \circ \iota$.


\end{document}